\documentclass{aa}
\input psfig.tex
\usepackage{graphicx}
\usepackage{natbib}
\usepackage{array}
\usepackage{graphics}
\usepackage{latexsym}
\usepackage{amssymb}
\usepackage{amsmath}
\usepackage{fancyhdr}
\usepackage{morefloats}
\usepackage{bm}
\bibpunct{(}{)}{;}{a}{}{,}

\begin{document}
\title{Environmental effects on star formation in \\
dwarf galaxies and star clusters}
\subtitle{}

\author{S.\ Pasetto\inst{1},
				M. Cropper\inst{1},
				Y. Fujita\inst{2},
				C. Chiosi\inst{3}
				\and
				E.K.\ Grebel\inst{4}
           }

\offprints{s.pasetto@ucl.ac.uk}

\institute{University College London, Department of Space \& Climate Physics, Mullard Space Science Laboratory, Holmbury St. Mary, Dorking Surrey RH5 6NT, United Kingdom
\and
Department of Earth and Space Science, Graduate School of Science, Osaka University,
Toyonaka, Osaka, Japan
\and
Physics and Astronomy Department, Padua University, Padua, Italy
\and
Astronomisches Rechen-Institut, Zentrum f\"ur Astronomie der Universit\"at Heidelberg, Heidelberg, Germany
}
\date{Received: /  Accepted 09 September 2014 }

\titlerunning{Environmental effects}
\authorrunning{S.\ Pasetto et al.}
\onecolumn

\abstract 
{The role of the environment in the formation of a stellar population is a difficult problem in astrophysics. The reason is that similar properties of a stellar population are found in star systems embedded in different environments or, vice versa, similar environments contain stellar systems with stellar populations having different properties.}
{In this paper we develop a simple analytical criterion to investigate the role of the environment on the onset of star formation.
We will consider the main external agents that influence the star formation (i.e. ram pressure, tidal interaction, Rayleigh-Taylor and Kelvin-Helmholtz instabilities) in a spherical galaxy moving through an external environment. The theoretical framework developed here has direct applications to the cases of dwarf galaxies in galaxy clusters and dwarf galaxies orbiting our Milky Way system, as well as any primordial gas-rich cluster of stars orbiting within its host galaxy.}
{We develop an analytic formalism to solve the fluid dynamics equations in a non-inertial reference frame mapped with spherical coordinates. The two-fluids instability at the interface between a stellar system and its surrounding hotter and less dense environment is related to the star formation processes through a set of differential equations. The solution presented here is quite general, allowing us to investigate most kinds of orbits allowed in a gravitationally bound system of stars in interaction with a major massive companion.}
{We present an analytical criterion to elucidate the dependence of star formation in a spherical stellar system (as a dwarf galaxy or a globular cluster) on its surrounding environment useful in theoretical interpretations of numerical results as well as observational applications. We show how spherical coordinates naturally enlighten the interpretation of the two-fluids instability in a geometry that directly applies to astrophysical case.  This criterion predicts the threshold value for the onset of star formation in a mass vs. size space for any orbit of interest. Moreover, we show for the first time the theoretical dependencies of the different instability phenomena acting on a system in a fully analytical way.} {}

\keywords{tidal forces, ram pressure, Rayleigh-Taylor, Kelvin-Helmholtz, dwarf galaxies, molecular clouds, star formation processes, stellar populations, colour magnitude diagrams }

\maketitle

\section{Introduction}\label{Introduction}

The effects of the environment on the evolution of a system are studied in several branches of physics, thermodynamics, statistical mechanics, and also astronomy. One such astronomical system is a galaxy. Galaxies are characterized by their large dimension and hence are dominated in their evolution by the gravitational force. Gravity is a long range force propagating at the speed of light and without a natural scale length. Thus from a theoretical point of view, every system close enough to a reference point (inside the cosmological horizon) is never fully isolated and represents a system in interaction with its environment. Examples of gravitational interactions available to date are the globular clusters inside the Milky Way (MW) \citep[e.g.,][]{1997ApJ...474..223G, 2001ApJ...548L.165O, 2010A&A...522A..71J}, the dwarf galaxies interacting around our MW \citep[e.g.,][]{2008PASA...25..121C, 2010ApJ...723.1618N, 2012ApJ...756...79S} and around the MW neighbour Andromeda \citep[e.g.,][]{2004ApJ...612L.117Z,2001Natur.412...49I,2006MNRAS.371.1983M, 2002AJ....124..310C}, and the closest groups of galaxies \citep[e.g.,][]{1994Natur.372..530Y,2002A&A...396..473M,2012A&A...541A.131C}.

A simple gravitational description of a galaxy would result in serious defect if it does not account for an appropriate description of its buildings blocks: the stars. The process of star formation is tightly connected with the gravitational evolution of a galaxy system. The interplay between star formation and gravitational evolution of a system has been extensively investigated in astronomy in the last century within the context of the Jeans instability \citep[][]{1902RSPTA.199....1J} passing through all its generalizations (as most recently in \cite{2013MNRAS.434L..56J}) or the star formation laws \citep[e.g.,][]{ 1959ApJ...129..243S}. The star formation regions are investigated both observationally \citep[e.g.,][]{ 2012MNRAS.422.3339W} and with numerical experiments \citep[e.g.,][]{2012ApJ...749..181F, 2012MNRAS.422.1609T}.

In a recent paper, \citet[][]{2012A&A...542A..17P} hereafter Paper I, the authors presented a technique to couple gravitational effects and star formation processes. The investigation of the role of external effects on star formation being the primary focus of that study, the authors developed a relation to express the pressure exerted by external phenomena on a primary system. In this way they were able to account for the roles of the external agents (e.g., an external hot gas, an external gravitational force etc.) on the system under examination. The standard Jeans instability criterion for stellar formation was substituted by a description ruled by a partial differential system of equations (PDEs) allowing them then to handle the molecular mass spectrum, as well as to obtain high mass resolution \citep[][]{1998ApJ...509..587F, 1999ApJ...516..619F}. 
In Paper I it was shown that it is possible to study the linear response of a gravitationally bound group of stars (e.g., a dwarf galaxy) in this way, and to capture the essence of what is observed in a dwarf galaxy like Carina during its interaction with the MW.

In this present work we take our theoretical investigation further. We account for the interaction between gravity and star formation by developing a new criterion of instability for the growth of the perturbation in an unstable fluid (molecular gas) where the star formation begins. 
This work is based on the seminal work by \citet{1954JAP....25...96P}, generalized to account for the non-inertial nature of the reference frame with the formalism presented in Paper I. We focus on the contrast between two gaseous systems of different density and temperature, such as the case of a gas-rich galaxy moving within a hot intra-cluster medium. In Plesset's work (but see also \citet{1958PhFl....1..201B}) the instability growth was followed in spherical coordinates for an expanding bubble. The instability condition was worked out and then extended in the following years to a Lagrangian description \citep[e.g.,][]{ 1978PhFl...21..140C}, to account for the viscosity of the medium \citep[e.g.,][]{ 1978PhFl...21.1465P} and for a stratified medium \citep[e.g.,][]{1990PhRvA..42.3400M}, etc. This theory is useful in various applications: in plasma physics, accelerated streams, Richtmyer-Meshkov instability, etc.

In our case, we generalize Plesset's technique to a non-inertial reference frame using the pressure equation derived in Paper I. We then apply the resulting equation to the case of the instability of two systems with a high density difference, as is the case for the hot intergalactic medium in a cluster of galaxies and the cold molecular clouds where stars form. 
The contents of the paper are the following: in Section \ref{LRT} the linear response theory is just introduced but formally developed in Appendix A. In Section \ref{Results} the resulting instability parameter is presented and explained. In Section \ref{Applicationandexamples} a few examples are illustrated. In Section 5 we summarize the results of the paper. Appendix A contains the full development of the theory representing the core of the paper: in Appendix \ref{Kinematicboundaryconditions} the kinematic boundary conditions between two fluids in relative motion are computation of the potential flow for internal (Appendix \ref{Internalgasperturbedpotentialflow}) and external (Appendix \ref{Externalgasperturbedpotentialflow}) gas. The dynamical boundary conditions are then evaluated in Appendix \ref{Internalgaspressureequation} for the internal gas pressure equation and in Appendix \ref{Externalgaspressureequation} for the external pressure equation. The equation for the surface of equilibrium is presented in Appendix \ref{Surfaceofequilibrium}. Finally the condition for instability is obtained in Appendix \ref{Conditionfortheinstability}. In Appendix B a few auxiliary functions defined in the text are analysed.

\section{Orbiting systems}\label{LRT}
\begin{figure*}
\sidecaption
\includegraphics[width=12cm]{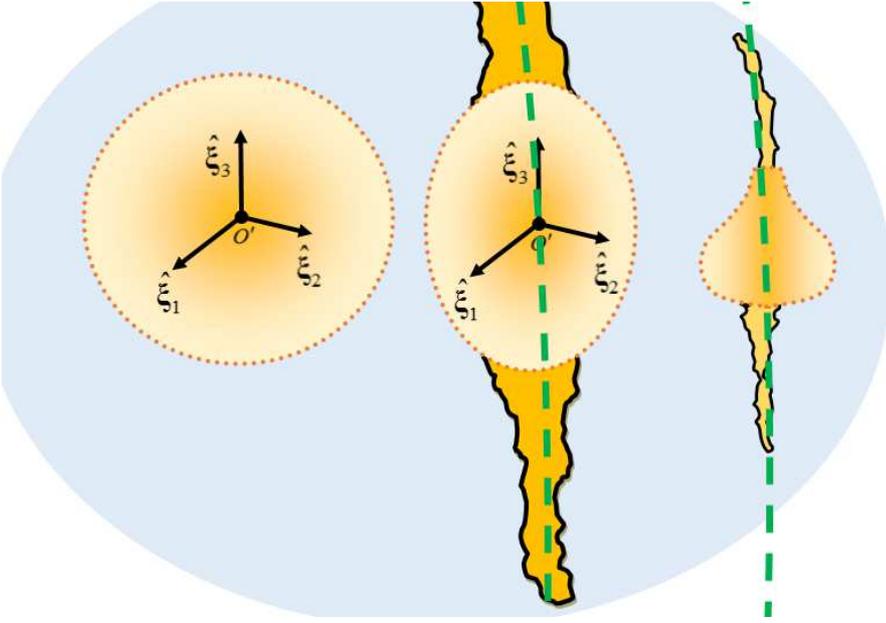}
\caption{$l=2$ perturbation mode of a spherical harmonics $Y_2^m\left( {\theta ,\phi } \right)$(central) over the unperturbed axisymmetric system (left). We exclude perturbation modes other than $l=2$ because, although every perturbed 3D surface can be realized by a superposition of spherical harmonics modes, no common evidence exists for the type of symmetries as the one presented on the right (e.g., $l=3$ mode). The dashed-green line provides an example of a star cluster orbit. Note that the tidal tails (yellow zone bordered by the black contours) do not necessarily lie along the orbits, i.e. ${\bm{O}}\left( t \right) \ne {\bm{1}}$ in the formalism of this paper, see e.g., \citet[][their Fig.7]{2010arXiv1009.2758P} or \citet{2004AJ....127.2753D,2005AJ....129.1906C} for globular cluster cases.}
\label{TTEvol}
\end{figure*}

The picture we are going to introduce is quite general, and suits several applications. Nevertheless, it is convenient to focus on a simple example. We consider two extended bodies consisting of a first system larger in mass and size, described by a density profile (or relative potential), and a secondary system smaller in mass such as a dwarf galaxy orbiting a major companion (e.g., MW dwarf galaxies or a spherical galaxy in a cluster of galaxies). 
We start considering a galaxy at rest or in a rectilinear motion, i.e. a single system not perturbed by external agents. We consider it to be well represented in the configuration space by a spherical geometry. Hence, despite its clumpy nature, we assume that the molecular gas, the site of the star formation, is well represented by a spherical distribution (in the literature the assumption of spherical geometry is extensively adopted from stellar clusters to clusters of galaxies). 
If we now consider this galaxy in interaction with external agents (tidal interaction with a perturbing system, ram pressure from external gas, etc.) its initial state of equilibrium in the velocity as well as in configuration space is perturbed (see Fig.\ref{TTEvol}).

In the following we are interested in quantifying the external effects acting on this galaxy gas distribution and on the star formation processes. 
The same treatment for the density profile perturbation of stars or dark matter can be achieved with the formalism developed in \citet[][]{1999ApJ...525..720C} or \citet[][]{1999MNRAS.306....1N} where star formation processes are nevertheless ignored. 
Here, we are going to neglect the internal mass distribution profile of the orbiting galaxy (or stellar cluster) by simply constructing the system with two parameters: mass $M$ and scale radius ${r_s}$. The internal gas component resulting mass distribution is for example given by $M\left( {{r_s}} \right) = \frac{4}{3}\pi r_s^3{\rho _{{\rm{gas}}}}$ and its gravitational radius by ${r_g} = \frac{5}{3}{r_s}$, despite any spherical couple potential-density can be considered (see Appendix A). The external major system description can be as complex as we like.

\subsection{Preliminaries: Internal processes and instabilities}
We start with the description of the surface of the galaxy in its motion throughout an intra-cluster medium (MW hot corona, galaxy cluster intergalactic medium etc.). We suppose that the galaxy, whose dimension we denote with ${r_s}$ (where ${r_s}$ can be thought to be the effective radius, the tidal radius, or any scale radius chosen for a particular purpose), is perturbed from its equilibrium as mentioned above. Then, the distribution of the molecular clouds in the interstellar medium of the galaxy (that we identify with the reservoir of gas for stellar formation) can be identified with a density distribution $\rho  = \rho \left( {\bm{\xi }} \right)$ bordered by a surface $\Sigma$ in a system of reference (SoR) comoving with the galaxy whose barycentric is in $O'$ and axis vectors $\hat \xi_i$ that we call $S_1=S_1(O',\hat \xi_i)$ for $i=1,2,3$ (for a more formal definition see Appendix A). This distribution is then perturbed to a new state, corresponding to a new perturbed surface density. Since we are interested in investigating only the instabilities ablating the gas from the stellar system or compressing it, we will limit ourselves to a linear analysis and we will assume the defining equation for the surface $\Sigma \left( {\xi ,\theta ,\phi ;t} \right)=0 $ (where spherical coordinates have been employed introduced in $S_1$) to reduce to
\begin{equation}\label{Eq02}
	\Sigma \left( {\xi ,\theta ,\phi ;t} \right) \equiv \xi  - \left({r_s}\left( t \right) + \eta \left( t \right)Y_l^m\left( {\theta ,\phi } \right)\right),
\end{equation}
because $\Sigma \left( {\bm{\xi }} \right)$ is defined by the value of the norm of the position vector $\xi  \equiv \left\| {\bf{\xi }} \right\| = {r_s}\left( t \right) + \eta \left( t \right)Y_l^m\left( {\theta ,\phi } \right)$ where ${r_s} = \left\| {{{\bm{r}}_s}} \right\|$, $\eta  \ll {r_s}$ is a real function (we omit its dependence on $l$  and $m$ ) and $Y_l^m = \sqrt {\frac{{2l + 1}}{{4\pi }}} \sqrt {\frac{{(l - m)!}}{{(l + m)!}}} {e^{im\phi }}\left( {P_l^m\left( \mu  \right)} \right)$ for $l \geqslant 0$  are the spherical harmonics with symmetry $Y_l^m\left( {\theta ,\phi } \right) = Y_{ - \left( {l + 1} \right)}^m\left( {\theta ,\phi } \right)$ for $l \leqslant  - 1$, and $P_l^m\left( \mu  \right)$ with $\mu  = \cos \theta $ the Legendre functions  \citep[e.g.,][]{1965PhT....18l..70L}. We are interested in the instability problem so we can omit the sum sign in Eq.\eqref{Eq02}  \citep[e.g.,][Chap. 1]{1961hhs..book.....C} and later on we will focus on the $l = 2$ perturbative mode. Nevertheless, in other to recover the correct literature flat-geometry limit we will keep $l$ unspecified for now. 
 (see Section \ref{Applicationandexamples}, and Fig\ref{TTEvol}). 

In Paper I, the authors established a framework to predict how a few selected instabilities (ram pressure, gas instabilities and tidal interactions) affect star formation. 
The key result of that work was a technique able to handle interacting systems in semi-analytical fashion. The authors obtained a pressure equation solution of the Navier-Stokes equations in a frame comoving with an orbiting stellar system. In this way, they were able to study the instabilities and the star formation through a pressure formulation within the scale radius of a system, ${r_s}$, and for a specific direction relative to the motion, that reduced to classical results of dimensionless galaxies \citep[e.g.,][]{1972ApJ...176....1G} as particular cases. Indeed, once an equation for the pressure $p$ was derived in a non-inertial reference frame, the star formation efficiency $\varepsilon \left( {{{\hat M}_i},p} \right)$ and lifetime of $\tau \left( {{{\hat M}_i},p} \right)$ of the mass spectrum of molecular clouds ${\hat M_i} \in \left[ {{{10}^2}{{,10}^6}} \right]{M_ \odot }$ was computed by following literature recipes \citep[e.g.,][]{1997ApJ...480..235E}. Here ${\hat M_i} \equiv {M_{i + 1}} - {M_i}$ defines the mass resolution with which the system of PDEs governing the evolution of the molecular clouds is integrated:
\begin{equation}\label{Eqsys01}
\frac{{d{{\hat M}_i}}}{{dt}} = \Xi \left( {{{\hat M}_i}} \right)\left( {{{\hat f}_i}\left( {{R_{{\rm{star}}}} + {R_{{\rm{mol}}}}} \right) - \frac{{{{\hat M}_i}}}{\tau }} \right),
\end{equation}
where ${{\hat f}_i} \equiv \frac{{{{\hat M}_i}}}{{{M_{{\mathop{\rm tot}\nolimits} }}}}$ with ${{M_{{\mathop{\rm tot}\nolimits} }}}$ being the total mass of the clouds, $\Xi \left( {{{\hat M}_i}} \right)$ is the step-function \citep[e.g.,][]{1972hmfw.book.....A}, ${R_{{\rm{star}}}} \equiv \int_{{m_{{\rm{low}}}}}^{{m_{{\rm{up}}}}} {\left( {\psi \left( {t - \tilde t} \right) - \psi \left( { - \tilde t} \right)} \right)\mu \left( m \right)\iota \left( m \right)dm} $ is the gas ejection rate from the stars. This gas ejection rate depends on the fraction of stars (of mass $m$ born at time ${\tilde t}$)  returned to the interstellar medium with return mass function $r(m)$ normalized  $\mu \left( m \right) \equiv \frac{{r\left( m \right)}}{m}$, on the stellar initial mass function $\iota \left( m \right)$, and on the star formation itself. The ${R_{{\rm{mol}}}} \equiv \sum\limits_i^{} {\left( {1 - \varepsilon } \right)\frac{{{{\hat M}_i}}}{\tau }} $ is the recycling rate of the molecular gas.
 Once this system is considered, an instability may gives rise to star formation if (and only if) gas is effectively present, i.e. $\Xi \left( {{{\hat M}_i}} \right)>0$, and suitable criteria dependent on the physics and geometry involved are met. In this case the resulting star formation is
\begin{equation}
\psi \left( t \right) = \sum\limits_i^{} {\varepsilon \left( {{M_i},p} \right)\frac{{{{\hat M}_i}}}{{\tau \left( {{M_i},p} \right)}}}. 	
\end{equation}
The star formation history can then be recovered once long-life, $m < 2.3{M_ \odot }$, and short-life $m \geqslant 2.3 M_\odot$ ($m_{\rm{low}}=0.08 M_\odot$ and $m_{\rm{up}}=100 M_\odot$) stellar feedback to the inter-stellar medium (ISM) is considered, following the recipe in \citet{1998ApJ...509..587F} and accounting for a two-phase ISM model (Field's instability) where a delay due to the HI phase is considered for the gas ejected by stars and evaporated by young stars  before it becomes finally molecular gas. 
Of course, this approach relies heavily on the stellar model adopted and on the time-scales of gas transitions. We followed the recipes depicted in \citet{1999ApJ...516..619F} combined with the stellar models of \citet{2009A&A...508..355B, 1995A&A...301..381B}. Different stellar models and ISM recipes can produce different time-scales for the remnants and as a consequence we consider our results as indicative only.

In this context the role of the instabilities was left to a description developed locally in a plane geometry approximation. The criteria adopted there were the standard literature instability conditions \citep[e.g.,][]{1961hhs..book.....C}. In particular, within the Paper I framework, the pressure on the molecular cloud of a dwarf galaxy was considered as a piston acting on a locally defined position of the dwarf - specified by the angle $\theta $ and radius ${r_s}$ - and there (i.e., determined “locally” for each point) the criterion for the growth of the instability was derived in the context  of the plane-geometry. The classical linear growth rate, $\gamma $, for the Rayleigh-Taylor (RT) and Kelvin-Helmholtz (KH) instabilities in a plane geometry can be derived by combining standard literature results \citep[e.g.,][]{1961hhs..book.....C} as:
\begin{equation}\label{Eq01}
	{\gamma ^2} = \frac{{{\rho _{{\text{out}}}}{\rho _{{\text{in}}}}{k^2}{{\left( {{v_{{\text{out}}}} - {v_{{\text{in}}}}} \right)}^2} + kg\left( {\rho _{{\text{out}}}^2 - \rho _{{\text{in}}}^2} \right)}}{{{{\left( {{\rho _{{\text{out}}}} + {\rho _{{\text{in}}}}} \right)}^2}}}.
\end{equation}
Here ${\rho _{{\text{out}}}}$ refers to the hot intergalactic medium external (outside) of  the galaxy (e.g., hot intra-cluster gas, MW hot coronal gas etc.), ${\rho _{{\text{in}}}}$ refers to the colder molecular cloud gas of the galaxy that will give rise (when unstable) to star formation processes, $k$ is the wave number of the instability, and $g$ the gravity acting on the system at the distance impacting the external pressure, $g = \frac{{GM}}{{{r_s^2}}}$ for unitary mass and mass $M$ at the distance ${r_s}$. If the fluid inside and outside an ideal surface of separation moves with relative velocity ${v_{{\text{rel}}}} \equiv {v_{{\text{out}}}} - {v_{{\text{in}}}} \ne 0$, then Eq.\eqref{Eq01} simultaneously accounts for the instability modes of sliding and pressing fluids, i.e. the KH or RT instabilities already considered in Paper I.

In this work, we will show how the description of the Paper I is simplified considerably in respect of the physical interpretation of the phenomena involved once the same instability growth criteria are followed directly in a spherical geometry. 
In order to achieve such a description, a few preliminary steps have to be performed in order to find a treatable reference frame for the equations involved. We start introducing the reference frame in the following section, a fundamental step to set the scene for the theory development and to understand our results. 

\subsection{Geometrical framework for potential flow approximation}\label{GeometricalSoR}
The framework follows closely that already introduced in \citet[][]{2009A&A...499..385P}. We consider the inertial reference frame attached to the more massive galaxy, ${S_0}$, and we call ${S_1}$ the \textit{reference system comoving with the smaller body}. In general, the axes of these two reference frames can be translated to match the same origin and overlapped by a rotation matrix ${\bm{O}} \in SO\left( 3 \right)$ with $\det  =  + 1$. Generally, if the smaller object is orbiting on its geodesic motion around the major one and ${S_1}$ is attached to it, this rotation matrix will be time dependent ${\bm{O}} = {\bm{O}}\left( t \right)$. This two-extended-body system will be considered in isolation.
The reader can visualize the abstract description of this paper if focusing on the image of a small stellar system, e.g., a dwarf galaxy or a globular cluster, centred in the origin of the system of reference $S_1$, and orbiting in the external potential of a cluster of galaxies or in the halo of the galaxies respectively.
As in Paper I, we will make use of the concept of the velocity potential. We assume the inter/intra-galactic medium to be irrotational \citep[e.g.,][Chap. 1]{1959flme.book.....L} $\nabla  \times {{\bm{v}}_0} = 0$ with ${{\bm{v}}_0}$ being the fluid velocity in ${S_0}$. Hence, there exists a scalar function ${\varphi _{{{\bm{v}}_0}}}$,  the velocity potential, whose gradient is the fluid velocity  i.e. $\exists {\varphi _{{{\bm{v}}_0}}}|{{\bm{v}}_0} = {\nabla _{\bm{x}}}{\varphi _{{{\bm{v}}_0}}}$. The ${\varphi _{{{\bm{v}}_0}}}$ is used in the Navier-Stokes equations to investigate the fluid dynamics of the two gas components: the one belonging to $S_1$'s galaxy and the one external to it. 
As in Paper I, we will make use of the concept of the velocity potential. We assume the inter/intra-galactic medium to be irrotational \citep[e.g.,][Chap. 1]{1959flme.book.....L} $\nabla  \times {{\bm{v}}_0} = 0$ with ${{\bm{v}}_0}$ being the fluid velocity in ${S_0}$. Hence, there exists a scalar function ${\varphi _{{{\bm{v}}_0}}}$,  the velocity potential, whose gradient is the fluid velocity  i.e. $\exists {\varphi _{{{\bm{v}}_0}}}|{{\bm{v}}_0} = {\nabla _{\bm{x}}}{\varphi _{{{\bm{v}}_0}}}$. The ${\varphi _{{{\bm{v}}_0}}}$ is used in the Navier-Stokes equations to investigate the fluid dynamics of the two gas components: the one belonging to $S_1$'s galaxy and the one external to it. 
However, as claimed in the introduction to this section, we are interested in providing a solution for the Navier-Stokes equation for the mentioned instabilities in a non-inertial reference frame. For this purpose, we have to picture the potential flow description of the Navier-Stokes equations solution in ${S_1}$ . 
\begin{figure*}
\sidecaption
\includegraphics[width=12cm]{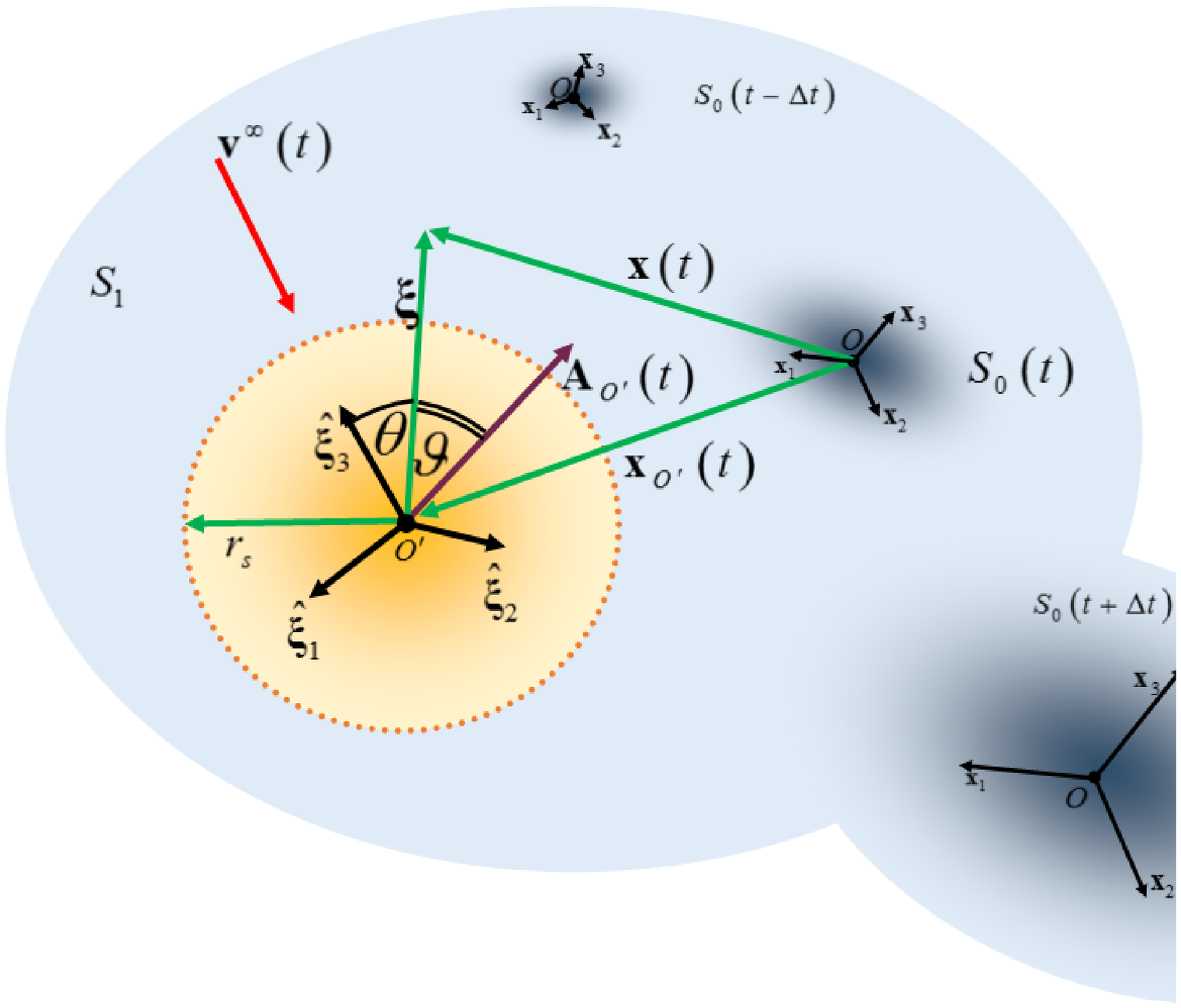}
\caption{Geometrical framework as seen by an observatory comoving with ${S_1}$. The position vector ${\bm{\xi }}$, the position vector of ${S_0}$, ${\bm{x}}\left( t \right)$ and the position of ${S_1}$ in ${S_1}$, ${{\bm{x}}_{O'}}\left( t \right)$, are shown in green. Here the system attached to the reference frame $S_1$ has been zoomed to show better the quantities defined in the text but it is supposed to be the smaller in mass and dimension and is orbiting around an inertial major system attached to $S_0$. The blue shadow represents the generic external environment attached to $S_0$-system in which the $S_1$-system is embedded. The position vector of ${S_0}$, once at the surface of the galaxy, $\Sigma $, take the sale radius value ${\left\| {\bm{\xi }} \right\|_{\Sigma}} = {r_s}$ (dot red orange circle). The scale radius is generally a function of time ${r_s} = {r_s}\left( t \right)$ and it varies as a consequence of the gravitational external field that the galaxy experiences  along its orbit around the major galaxy centred in ${S_0}$. 
In the non-inertial reference frame ${S_1}$  the velocity $v = v\left( t \right)$  (red arrow aligned with $\bm{\hat \xi }_3$) of the fluid impacting the galaxy (i.e. the negative of the velocity of the stellar system in ${S_0}$) forms an angle $\theta $ with the position vector ${\bm{\xi }}$, $\theta \equiv \widehat {\left({\bm{\xi }},{\bm{v}}\right)}$ (with $\widehat {\left({*},{*}\right)}$ notation for the smaller angle between two three-dimensional vectors). The acceleration (violet arrow)– inclusive of the apparent effect forced due to the non-inertial character of ${S_1}$ forms an angle $\vartheta  = \widehat {\left({\bm{\xi }},{{\bm{a}}_{O'}}\right)}$ with the position vector. An observer comoving with ${S_1}$  sees the inertial reference frame ${S_0}$ changing position at  different times $t - \Delta t$,$t$  or $t + \Delta t$ as well as the rotation of $S_0$ axes.}
\label{SoR}
\end{figure*}

The geometry of the problem is as shown in Fig. \ref{SoR}, where ${\bm{\xi }}$ is the arbitrary but fixed position vector in the SoR $S_1$ introduced above: ${S_1}\left( {O';{{{\bm{\hat \xi }}}_1},{{{\bm{\hat \xi }}}_2},{{{\bm{\hat \xi }}}_3}} \right)$ centred $O'$, with unitary vectors ${{\bm{\hat \xi }}_i}$ $i = 1,2,3$; ${\bm{x}} = {\bm{x}}\left( t \right)$ the position vector in ${S_0}\left( {O\left( t \right),{{{\bm{\hat x}}}_1}\left( t \right),{{{\bm{\hat x}}}_2}\left( t \right),{{{\bm{\hat x}}}_3}\left( t \right)} \right)$ centred in $O$ whose orbit as seen by an observer in $O'$ is $O = O\left( t \right)$; where ${{\bm{x}}_{O'}} = {{\bm{x}}_{O'}}\left( t \right)$ the position of the ${S_1}$ origin in ${S_0}$. 
The external velocity potential fluid was introduced in Paper I: the potential flow past a spheroidal dwarf galaxy is approximated by a classical literature result, $\varphi _{{{\bm{v}}_0}}^I \equiv \frac{1}{2}\left\langle {{{\bm{v}}^\infty },{\bm{x}}} \right\rangle \frac{{r_s^3}}{{{{\left\| {\bm{x}} \right\|}^3}}}$ that gives the potential flow in ${S_1}$ when added to a translational potential flow $\varphi _{{{\bm{v}}_0}}^{II} \equiv \left\langle {{{\bm{v}}^\infty },{\bm{x}}} \right\rangle $ that ``brings the galaxy to rest''. ${\varphi _{{{\bm{v}}_1}}} \equiv \varphi _{{{\bm{v}}_1}}^I + \varphi _{{{\bm{v}}_1}}^{II} =  - \left\langle {{\bm{v}},{\bm{\xi }}} \right\rangle \left( {1 + \frac{1}{2}\frac{{\xi _s^3}}{{{{\left\| {\bm{\xi }} \right\|}^3}}}} \right)$, thanks to the scalar character of the velocity potential (we recall that ${\bm{v}}^\infty$ is the velocity of the fluid at infinity and ${\bm{v}}$ the velocity of the $S_1$ system, $v = \left\| {\bf{v}} \right\|$ the speed obtained with standard Euclidean norm $\left\| * \right\|$, $\left\langle {*,*} \right\rangle$ the standard inner product between two vectors).

The description of the motion in ${S_1}$ instead of ${S_0}$ has some advantages in the mathematical treatment of the fluid dynamics equations. This is not a new approach to the Navier-Stokes equation and represents a standard literature procedure when dealing with two-fluid problems \citep[e.g.,][]{2000ifd..book.....B, 1959flme.book.....L}. In this way it is possible to simplify the description of the two-fluid interaction to a common reference frame: it is simple to prove that if the fluid is irrotational in a given reference frame it is not in another, being the vorticity, say ${\bm{\zeta }}$, a concept relative to the reference frame as the velocity (${{\bm{\zeta }}_0} = {{\bm{\zeta }}_1} + 2{\bm{\Omega }}$, with ${\bm{\Omega }}$ relative rotational velocity of ${S_1}$  and ${S_0}$ where the vorticity is called ${{\bm{\zeta }}_1}$  and ${{\bm{\zeta }}_0}$  respectively). The description of the motion in a non-inertial reference frame simplifies this approach. We also follow standard literature results in formulating the potential flow in relation to the velocity of the stellar system ${\bm{v}}$ instead of the velocity of the impacting flow ${{\bm{v}}^\infty }$ which simplifies the physical interpretation of our results. Finally, with ${\varphi _{{{\bm{v}}_1}}^{III}} \equiv  - \frac{{{{\dot r}_s}r_s^2}}{{\left\| {\bm{\xi }} \right\|}}$ \citep[e.g.,][]{2000ifd..book.....B}, we describe the potential flow of the gas internal to the galaxy alone. Moreover, we add it to the description of the external flow impacting the galaxy's internal molecular cloud gas when necessary (for example to describe the hot MW coronal gas).

The description presented so far was initially introduced in Paper I. However it has some limitations and imprecisions that do not permit the best understanding of the involved physics. Despite its success in reproducing the star formation history of the Carina dwarf galaxy presented in Paper I, the formalism there developed did not account properly for the deformation of the dwarf galaxy because of tidal interaction, and hence for the star formation instability presented in a real system. In any physical case, we  expect the system to suffer a geometric compression in a direction roughly orthogonal to the orbit and an elongation along the orbits where the tidal tails lie \citep[e.g.,][]{2003A&A...405..931P, 2011A&A...525A..99P}. Vice versa, we expect that elongation to be tilted by about $\frac{\pi }{2}$, with respect to the orbital direction proximate to the pericentre passages \citep[e.g.,][]{2009MNRAS.400.2162K}. In both the extreme cases, we want to be able to follow the impact of the pressure on a galaxy foliated by homoeoidal surfaces tilted with an arbitrary rotation matrix ${\bm{O}}$ introduced above (see Fig.\ref{TTEvol}).

In order to achieve such a generalization and simultaneously to investigate the role of the star formation instability, in Appendix A we will develop a two-fluids instability analysis in spherical geometry for a non-inertial reference frame under the influence of a non-uniform external gravitational field. The development of the theory proceeds as in the plane geometrical case of Paper I with two additional difficulties:
\begin{itemize}
	\item the presence of apparent forces owing to the non-inertial nature of the geometrical frame we used; 
	\item the deformation from spherical to oblate-spheroidal to address the limitation of the simple spherical geometry in the description of the tidal interaction of a system with an external gravitational field.
\end{itemize}
The linear response theory is described separately in Appendix A in order to give space here to the results and applications. Our key result take the form of an instability criterion that can be evaluated once a small parameter space is considered for the stellar system and its environment. The result is inspired by the standard quantum-mechanics Wenntzel-Kramers-Brillouin (WKB) approximation for the solution of evolution equations with slowly varying coefficients, but limited to the analysis of the condition on the positivity of the growth factor ${\gamma ^2}\left( {\theta} \right) > 0$ of the perturbation of a stellar system in motion. The result is obtained in Eq.\eqref{ApEq33} of Appendix A. After some algebra by collecting properly the terms and the trigonometric functions, this can be written as
\begin{align}\label{Eq36_2}
 {\gamma ^2} &= \frac{3}{2}\frac{{v_{\rm{rel}}\cos \theta }}{{r_s^2}}\left( {A + 1} \right)\left( {\frac{3}{2}v_{\rm{rel}}\cos \theta \left( {A + 1} \right) + {{\dot r}_s}} \right)\nonumber\\
&+ \frac{9}{4}\frac{{\left( {A - 2} \right)\left( {A + 1} \right){F_1}{v_{\rm{rel}}^2}\sin \theta \cos \theta }}{{2r_s^2}}\nonumber\\
 &+ \frac{{a_{{\rm{O'}}}^ \bot }}{{{r_s}}}\left( {{\rm A}\left( {l - \frac{1}{4}} \right) - \frac{1}{4}} \right)\nonumber\\
 &+ \frac{9}{4}{v_{\rm{rel}}^2}{\sin ^2}\theta {\left( {\frac{{A + 1}}{{2{r_s}}}} \right)^2}\left( {F_1^2 - 2\frac{{{l_ - } + {F_2}}}{{A + 1}}} \right)\nonumber\\
&+ \left( {l + \frac{1}{2}} \right)A\frac{{{{\ddot r}_s}}}{{{r_s}}} + \frac{3}{4}\frac{{\dot r_s^2}}{{r_s^2}},
\end{align}
where we introduce the generalized Atwood number\footnote{Note how our definition differs from Eq. (14) in \citet[][]{1954JAP....25...96P} (and the form widely used in literature) because in the original definition the dependence of ${\rm A} = {\rm A}\left( {l,{\rho _{{\rm{in}}}},{\rho _{{\rm{out}}}},{r_s},{{\ddot r}_s}} \right)$) while we prefer to keep the original dimensionless nature of the Atwood number.}: 
\begin{equation}\label{Eq29}
	{\rm A} \equiv \frac{{{l_ + }{l_{ +  + }}{\rho _{{\rm{in}}}} - {l_ - }l{\rho _{{\rm{out}}}}}}{{{l_ + }{\rho _{{\rm{in}}}} + l{\rho _{{\rm{out}}}}}},
\end{equation}
Here $\theta $ is angle between the position vector ${\bf{\xi }}$ and the stellar system velocity vector in ${S_0}$. ${a_{O'}}$ and ${v_{{\rm{rel}}}}$ are the relative acceleration and velocity of ${S_1}$ in ${S_0}$, ${r_s}\left( t \right)$ is the selected scale radius of the system (with its velocity ${\dot r_s}$  and acceleration ${\ddot r_s}$), and finally ${\rho _{{\rm{in}}}}$  and ${\rho _{{\rm{out}}}}$  are the gas density inside (HI, molecular) or outside (hot interstellar medium) of the system being examined. The special functions ${F_1}$  and ${F_2}$  are auxiliary functions defined in Appendix B. The spherical harmonic azimuthal modes $l$ (with $l_+\equiv l+1$, $l_{++}\equiv l+2$ etc.) are used to account for the departure of the tidal deformation of the stellar system from its starting spherical shape (see Fig.\ref{TTEvol}).  Finally, because of the short life-time of the molecular clouds compared with the orbital time of the stellar systems considered, we can safely assume ${\rho _{{\rm{out}}}}$ and ${\rho _{{\rm{in}}}}$ - and ultimately the Atwood number - to be in local thermodynamic equilibrium. This will result in a further simplification of our equations described in the following sections.

\section{Results}\label{Results}
To interpret the role of the stability phenomena in the evolution of a stellar system centred on $S_1$ orbiting a major companion centred on $S_0$, it is convenient to make a few minor changes to Eq.\eqref{Eq36_2}. 
We split the velocity components of the external fluid into a parallel $v_{{\rm{rel}}}^ \bot  \equiv {v_{{\rm{rel}}}}\cos \theta $  and a perpendicular $v_{{\rm{rel}}}^\parallel  \equiv {v_{{\rm{rel}}}}\sin \theta $ component to the position vector in ${S_1}$. The same is done for the acceleration: with $\vartheta $ instead of $\theta$ we will proceed to define $a_{{\rm{O'}}}^ \parallel$ and $a_{{\rm{O'}}}^ \bot$. 
Moreover, while in Section \ref{Dynamicalboundarycondition} the algebra is laid out with the generic $l$ to prove that we are able to recover the plane limit in Section \ref{SpLims}, here only the $l = 2$ perturbation case is of interest. With Eq.\eqref{Eq29} for $l=2$, we can rewrite the growth factor as:
\begin{align}\label{Eq37}
\gamma _{l = 2}^2 \equiv \hat \gamma^2 &= \frac{5}{2}\hat {\rm A}\frac{{{{\ddot r}_s}}}{{{r_s}}} + \frac{3}{4}\frac{{\dot r_s^2}}{{r_s^2}}\nonumber\\
 &+ \frac{3}{2}\frac{{v_{{\rm{rel}}}^ \bot {{\dot r}_s}}}{{r_s^2}}\left( {\hat {\rm A} + 1} \right) + \frac{9}{{16}}\frac{{v_{{\rm{rel}}}^{ \bot 2}}}{{r_s^2}}{\left( {\hat {\rm A} + 1} \right)^2}\nonumber\\
 &+ \frac{9}{{16}}\frac{{v_{{\rm{rel}}}^{\parallel 2}}}{{r_s^2}}\left( {\hat {\rm A} + 1} \right)\left( {\left( {\hat {\rm A} + 1} \right)\hat F_1^2 - 2\left( {{{\hat F}_2} + 1} \right)} \right)\nonumber\\
 &+ \frac{9}{{16}}\frac{{v_{{\rm{rel}}}^\parallel v_{{\rm{rel}}}^ \bot }}{{r_s^2}}\left( {\hat {\rm A} - 2} \right)\left( {\hat {\rm A} + 1} \right){{\hat F}_1}\nonumber\\
 &+ \frac{{a_{O'}^ \bot }}{{4{r_s}}}\left( {7\hat {\rm A} - 1} \right),
\end{align}
with
\begin{equation}
{{\rm A}_{l = 2}} \equiv \hat {\rm A} = \frac{{2{\rho _{{\rm{out}}}} - 3{\rho _{{\rm{in}}}}}}{{2{\rho _{{\rm{out}}}} + 3{\rho _{{\rm{in}}}}}}
\end{equation}
where we additionally defined the special function ${\hat F_1} \equiv {F_1}|_{l = 2}$ and ${\hat F_2} \equiv {F_2}|_{l = 2}$.
This is the fundamental result of this paper, and it gives us the interpretation key of the stability phenomena globally acting on a spherical stellar system. To first order the growth of the instabilities is the sum of different contributions, 
\begin{equation}\label{Gamma2parameter}
{\hat \gamma ^2} = \hat \gamma _{\rm{I}}^2 + \hat \gamma _{{\rm{RT}}}^2 + \hat \gamma _{{\rm{KH}}}^2 + \hat \gamma _{{\rm{mix}}}^2 + \hat \gamma _{{\rm{a - RT}}}^2,	
\end{equation}
which we identify line by line:
\begin{enumerate}
	\item The terms in the first line of Eq.\eqref{Eq37}
	\begin{equation}\label{Eq38}
		\hat \gamma^2_\text{I} \equiv \frac{5}{2}\hat {\rm A}\frac{{{{\ddot r}_s}}}{{{r_s}}} + \frac{3}{4}\frac{{\dot r_s^2}}{{r_s^2}}=\frac{{5\left( {2{\rho _{{\rm{out}}}} - 3{\rho _{{\rm{in}}}}} \right){{\ddot r}_s}}}{{2{r_s}\left( {3{\rho _{{\rm{in}}}} + 2{\rho _{{\rm{out}}}}} \right)}} + \frac{{3\dot r_s^2}}{{4r_s^2}},
	\end{equation}
	which is the standard literature result for an inertial reference frame. These terms do not depend on the subject of our study: for example they are in common with previous studies on the growth of bubbles in an inertial reference frame or supernova explosions. Here, though, we limit ourselves to the $l = 2$ mode of disturbance because of the physical system under consideration. This mode contributes positively to the instability every time  $\frac{3}{2}\frac{{\dot r_s^2}}{{{r_s}}}\left( {3{\rho _{{\rm{in}}}} + 2{\rho _{{\rm{out}}}}} \right) > \frac{5}{4}\left( {3{\rho _{{\rm{in}}}} - 2{\rho _{{\rm{out}}}}} \right){\ddot r_s}$, and when assuming $3{\rho _{{\rm{in}}}} - 2{\rho _{{\rm{out}}}} > 0$ in the case of galaxies or stellar clusters moving through a hot medium. Its positivity depends for example on the expansion/contraction of the tidal radius of the stellar system. Equation \eqref{Eq37}, if solved together with the corresponding Eq.\eqref{Eq26} of the equilibrium surface, will eventually lead to the complete eigenvalues of the system which can directly be compared with numerical simulation;
	\item	The terms in the next line of Eq.\eqref{Eq37}
	\begin{equation}\label{Eq39}
		\hat \gamma^2_{\text{RT}} \equiv \frac{3}{2}\frac{{v_{{\rm{rel}}}^ \bot {{\dot r}_s}}}{{r_s^2}}\left( {\hat {\rm A} + 1} \right) + \frac{9}{{16}}\frac{{v_{{\rm{rel}}}^{ \bot 2}}}{{r_s^2}}{\left( {\hat {\rm A} + 1} \right)^2}=\frac{{9\rho _{{\rm{out}}}^2v_{{\rm{rel}}}^{ \bot 2}}}{{r_s^2{{\left( {3{\rho _{{\rm{in}}}} + 2{\rho _{{\rm{out}}}}} \right)}^2}}} + \frac{{6{\rho _{{\rm{out}}}}{{\dot r}_s}v_{{\rm{rel}}}^ \bot }}{{r_s^2\left( {3{\rho _{{\rm{in}}}} + 2{\rho _{{\rm{out}}}}} \right)}},
	\end{equation}
	 proportional to $v_{{\rm{rel}}}^ {\bot 2} $. We will call these terms ``pure''-RT terms. They influence the instability owing to the pressure along the radial extension of the star cluster. They show how the instability dependence on the RT effect is quadratic on the velocity of the fluid impacting the galaxy, i.e. quadratic on the velocity of the stellar system itself. It is especially interesting to observe how the term $\propto v_{{\rm{rel}}}^{ \bot 2}$ is always present: even if ${\dot r_s} = 0$, for example in the case of a galaxy that has reached its equilibrium by violent relaxation, the instability grows with quadratic dependence on the orbital velocity. As expected, it is maximum at the stagnation point, and it decreases slowly away from the direction of motion, becoming formally zero at $\theta  = \frac{\pi }{2}$;
	\item The term 
	\begin{equation}\label{Eq40}
		\hat \gamma^2_{\text{KH}} \equiv \frac{9}{{16}}\frac{{v_{{\rm{rel}}}^{\parallel 2}}}{{r_s^2}}\left( {\hat {\rm A} + 1} \right)\left( {\left( {\hat {\rm A} + 1} \right)\hat F_1^2 - 2\left( {{{\hat F}_2} + 1} \right)} \right)=\frac{{9{\rho _{{\rm{out}}}}v_{{\rm{rel}}}^{\parallel 2}\left( {2{\rho _{{\rm{out}}}}\hat F_1^2 - \left( {3{\rho _{{\rm{in}}}} + 2{\rho _{{\rm{out}}}}} \right)\left( {{{\hat F}_2} + 1} \right)} \right)}}{{2r_s^2{{\left( {3{\rho _{{\rm{in}}}} + 2{\rho _{{\rm{out}}}}} \right)}^2}}},
	\end{equation}
 proportional to $v_{{\rm{rel}}}^{\parallel 2}$.	We will refer to this term as the ``pure''-KH term. This is influenced by the sliding of the relative velocity between the dwarf galaxy ISM and the inter-cluster medium through which the stellar system is moving. As in the RT case, $\hat \gamma^2_{\text{KH}}$ is quadratic with the velocity of motion of the stellar system and does not depend on the radial expansion or contraction of the galaxy. Further insight in the understanding of this term will be gained in the next section where the astrophysical case of interest ${\rho _{{\rm{in}}}} \gg {\rho _{{\rm{out}}}}$ will be developed; 
\item The term 
\begin{equation}\label{Eq41}
	\hat \gamma^2_{\text{mix}} \equiv -\frac{9}{{16}}\frac{{v_{{\rm{rel}}}^\parallel v_{{\rm{rel}}}^ \bot }}{{r_s^2}}\left( {\hat {\rm A} - 2} \right)\left( {\hat {\rm A} + 1} \right){{\hat S}_1}=- \frac{{9{\rho _{{\rm{out}}}}v_{{\rm{rel}}}^ \bot v_{{\rm{rel}}}^\parallel \left( {9{\rho _{{\rm{in}}}} + 2{\rho _{{\rm{out}}}}} \right){{\hat F}_1}}}{{2r_s^2{{\left( {3{\rho _{{\rm{in}}}} + 2{\rho _{{\rm{out}}}}} \right)}^2}}}.
\end{equation}
This is a mixed-contribution term. It always exists except for the special case of the stagnation point or the tangential point where it disappears, either $v_{{\rm{rel}}}^ \bot $ or $v_{{\rm{rel}}}^\parallel $ being null, respectively. It is a quadratic term in the velocities, and it shows how the coexistence of KH and RT instabilities is always present once the galaxy is in motion along its geodesic. Its contribution to the instability depends on the sign of ${\hat F_1}$, being negative for small angles, contributing positively to the growth of the instability near the stagnation point, and positive for $\theta  \simeq \frac{\pi }{2}$ thus having a stabilizing factor against the “pure”-KH term introduced before (See Fig A.1 in Appendix A);
\item The term 
\begin{equation}\label{Eq42}
	\hat \gamma^2_{\text{a-RT}} \equiv \frac{{a_{\rm{O'}}^ \bot }}{{4{r_s}}}\left( {7\hat {\rm A} - 1} \right)=\frac{{a_{{\rm{O'}}}^ \bot \left( {3{\rho _{{\rm{out}}}} - 6{\rho _{{\rm{in}}}}} \right)}}{{{r_s}\left( {3{\rho _{{\rm{in}}}} + 2{\rho _{{\rm{out}}}}} \right)}},
\end{equation}
 proportional to the acceleration component in the direction indicated by the position vector. This term is  a completely new result of our theory (it cancels out at the plane geometry limit). This term has a different nature from the terms originally described in the works of Kelvin, Helmholtz, Rayleigh and Taylor: this term is an apparent \textit{force }due to the non-inertial nature of the reference system we adopted. It shows a linear dependence on the acceleration $a_{{\rm{O'}}}^ \bot$ to contribute orthogonally to the surface of the galaxy, i.e., only along its radial direction. Hence, to the first order, this term contributes (with a positive or negative force, stabilizing or promoting instability) only to the RT instability, not to the KH one. Clearly this term has a different contribution to the instability depending on the actual orbit and on the angle $\vartheta $.
\end{enumerate}
Curiously, \textit{our results indicate no direct contribution of the tangential component of the acceleration to the overall instability $a_{{\rm{O'}}}^\parallel$, i.e. we discovered that no apparent force acts on the KH type of instability to the first order}. This component has nevertheless to be present at second order, as evident in the equilibrium equation derived in Appendix A Eq.\eqref{Eq26} to the leading order, or in the growth factor Eq.\eqref{Eq37} when analysed to the second order. We mention that  in the more complicated work by \citet{2006PhFl...18g2104S} a similar analysis to ours is carried out to higher order but for a non-translational system of reference. We remark that even besides the technical difficulties in carrying out such an analysis in our non-inertial case, this is not of interest in our case: in Paper I we showed that the life-time of the molecular clouds subject to external pressure is below 300 Myrs for the Local Group case. Hence, within these timescales, higher order terms or resonances do not have time to play a role.
A general treatment of the force acting on the generic point of an element in ${S_1}$ was given by Eq.(6) of \citet{2009A&A...499..385P} that in tidal approximation reads ${{\bm{\ddot r}}_s} = {{\bm{O}}^{\rm{T}}}{\bm{TO}}{{\bm{r}}_s} - 2{\bm{\Omega }} \times {{\bm{\dot r}}_s} - {\bm{\dot \Omega }} \times {{\bm{r}}_s} - {\bm{\Omega }} \times \left( {{\bm{\Omega }} \times {{\bm{r}}_s}} \right)$. Eq.\eqref{Eq36_2} holds only in the case of the short lifetime of the dwarf galaxy's molecular clouds we are considering (see Fig.(1) in \citet{2012A&A...542A..17P}). 

\section{Application and examples}\label{Applicationandexamples}
In what follows, we develop some analytical, numerical and theoretical examples and exercises to show the potential of the criterion developed above.

\subsection{Instability for the case ${\rho _{{\rm{out}}}} \ll {\rho _{{\rm{in}}}}$}
To gain better insight into the physical conditions for the positivity of the growth factor, we consider the special case where the hot inter-galactic medium, here ${\rho _{{\rm{out}}}}$, is much more diffuse than the cold molecular clouds density distribution, ${\rho _{{\rm{in}}}}$, that we are considering as the star formation site. In this case a lighter fluid is pressing on a heavier one described in the non-inertial reference frame ${S_1}$. This is a practical case of interest in astrophysics. Because the density difference between the hot intergalactic medium and molecular clouds is assumed to be extremely high,  ${\rho _{{\rm{in}}}} \gg {\rho _{{\rm{out}}}}$ and we can expand the previous Eq. \eqref{Eq37} to get to the first order in the small parameter $\varepsilon  = \frac{{{\rho _{{\rm{out}}}}}}{{{\rho _{{\rm{in}}}}}}$:
\begin{equation}\label{Eq43}
	{\hat \gamma ^2} =  - \frac{{9\varepsilon v_{{\rm{rel}}}^ \bot v_{{\rm{rel}}}^\parallel {{\hat F}_1}}}{{2r_s^2}} - \frac{{3\varepsilon v_{{\rm{rel}}}^{\parallel 2}({{\hat F}_2} + 1)}}{{2r_s^2}} + \frac{{(7\varepsilon  - 6)a_{O'}^ \bot }}{{3{r_s}}} + \frac{{2\varepsilon {{\dot r}_s}v_{{\rm{rel}}}^ \bot }}{{r_s^2}} + \frac{{10(4\varepsilon  - 3){r_s}{{\ddot r}_s} + 9\dot r_s^2}}{{12r_s^2}},
\end{equation}
whose positivity, for example at the stagnation point is simply:
\begin{align}\label{Eq44}
&{\hat \gamma ^2} = \frac{{(7\varepsilon  - 6)a_{O'}^ \bot }}{{3{r_s}}} + \frac{{2\varepsilon {{\dot r}_s}v_{{\rm{rel}}}^ \bot }}{{r_s^2}} + \frac{{10(4\varepsilon  - 3){r_s}{{\ddot r}_s} + 9\dot r_s^2}}{{12r_s^2}} > 0 \Leftrightarrow \nonumber\\
&\frac{{7\varepsilon  - 6}}{3}a_{O'}^ \bot  + \frac{{5\left( {4\varepsilon  - 3} \right)}}{6}{{\ddot r}_s} >  - 2\varepsilon \frac{{{{\dot r}_s}v_{{\rm{rel}}}^ \bot }}{{{r_s}}} - \frac{3}{4}\frac{{\dot r_s^2}}{{{r_s}}},
\end{align}
which shows a competition between the relative acceleration of the two reference frames ${S_1}$  and ${S_0}$, the gravity of the systems ${\ddot r_s} = g =  \frac{{GM}}{{r_s^2}}$, the velocity terms $v_{{\rm{rel}}}^ \bot$  and the contraction velocity ${\dot r_s}$. At the limit of $\varepsilon  \to 0$ there are no hydrodynamical effects and the gas instability will be purely gravitational. We get:
\begin{equation}\label{Eq45}
	2a_{O'}^ \bot  + \frac{5}{2}g < \frac{3}{4}\frac{{\dot r_s^2}}{{{r_s}}},
\end{equation}
satisfied in the zones of the galaxy where $a_{O'}^ \bot  < 0$, i.e. where the component of the external acceleration compresses the gas. This is indeed a well known literature result on the purely gravitational compressive effect of a tidal field acting on a galaxy, that we recover with our stability criteria. The dissipative phenomena in the pure dynamical case are still a matter of debate \citep[e.g.,][]{2014arXiv1406.2376E,2013MNRAS.434L..56J} that we avoid here. We simply limit ourself to observe that with the total potential acting at the point of interest on $\Sigma$ as 
\begin{align}\label{Eq47}
{\Phi _{{\rm{cl}}}}\left( {\bm{x}_\Sigma} \right) &\simeq {\Phi _{{\rm{cl}}}}\left( {{{\bm{x}}_{O'}}} \right) + {\partial _{\bm{x}_\Sigma}}{\Phi _{{\rm{cl}}}}\left( {{{\bm{x}}_{O'}}} \right)\left( {{\bm{x}_\Sigma} - {{\bm{x}}_{O'}}} \right) + ...\nonumber\\
{\partial _{\bm{x}_\Sigma}}{\Phi _{{\rm{cl}}}}\left( {\bm{x}_\Sigma} \right) &\simeq {\partial _{\bm{x}_\Sigma}}{\Phi _{{\rm{cl}}}}\left( {{{\bm{x}}_{O'}}} \right) + \partial _{\bm{x}_\Sigma}^2{\Phi _{{\rm{cl}}}}\left( {{{\bm{x}}_{O'}}} \right)\left( {{\bm{x}_\Sigma} - {{\bm{x}}_{O'}}} \right) + ...,
\end{align}
so that
\begin{align}\label{Eq48}
a_{{{O'}}}^ \bot &=\left\langle { - {\partial _{\bm{x}_\Sigma}}{\Phi _{{\rm{cl}}}}\left( {\bm{x}_\Sigma} \right),{\bm{O\xi }}} \right\rangle \nonumber\\ 
&\simeq \left\langle { - {\partial _{\bm{x}_\Sigma}}{\Phi _{{\rm{cl}}}}\left( {{{\bm{x}}_{O'}}} \right),{\bm{O\xi }}} \right\rangle  - \left\langle {\partial _{\bm{x}_\Sigma}^2{\Phi _{{\rm{cl}}}}\left( {{{\bm{x}}_{O'}}} \right){\bm{O\xi }},{\bm{O\xi }}} \right\rangle \nonumber\\
 &= \left\langle {{{\bm{a}}_{O'}},{\bm{O\xi }}} \right\rangle  + \left\langle {{{\bm{O}}^{\rm{T}}}{\bm{TO\xi }},{\bm{\xi }}} \right\rangle, 
\end{align}
to which we want to add the stellar cluster mass distribution at the same position ${g} = \frac{{GM}}{{r_s^2}}$. This proof follows tightly the derivation of Eqs. 9 and 10 of Paper I and holds only for small systems orbiting major companions. 

\begin{figure*}
\sidecaption
\includegraphics[width=12cm]{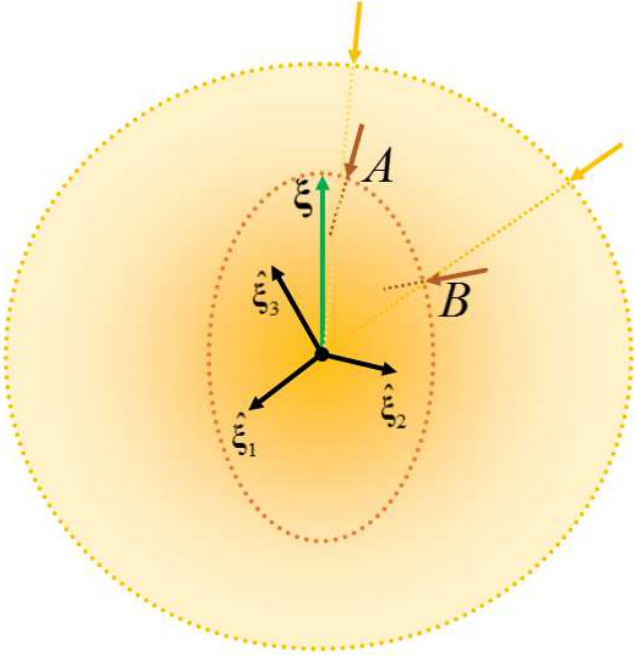}
\caption{Pictorial representation of small-angle interpretation. Formally our request for a small perturbation $\eta  \ll {r_s}$ should induce a small-angles interpretation of our results. Nevertheless, we see graphically the reason why we should expect our approximation to hold even if $\eta$ is not small. As evident in this figure, around the point A, we see the orthogonal direction to the equilibrium surface (dotted yellow) and to the perturbed surface (dotted orange) remains approximatively collinear. Vice versa, for larger angles (see point B in figure) the approximation is less good (and the functions $F_1$, $F_2$ present a divergence, see Appendix A). We recover fully the validity of our approximation to the orthogonal case $\theta  \sim \frac{\pi }{2}$. Note how this divergence can be cured with higher order expansions in $\eta$.}
\label{PiccoliAngoli}
\end{figure*}

\subsection{Small angles $\theta  \cong 0$}
We are obviously interested in the small angles approximation. This is because in the spherical geometry that we have developed, the stagnation point lies where the pressure is higher, i.e. it is the first point impacting on the external intra-galaxy medium. Vice versa in a different geometry this is not necessarily true. If we assume a spiral galaxy penetrating a cluster of galaxies with a hot intra-cluster medium in a direction orthogonal to the disk plane, the first instability to be seen is the stripping from the border of the disk because of the weaker galaxy potential at the edges of the disk \citep[e.g.,][]{ 2008A&A...483..121R, 2005A&A...433..875R}. The instability criterion of Eq.\eqref{Eq37} reduces to 
\begin{equation}\label{Eq46}
	{\hat \gamma ^2} \simeq \frac{{9\rho _{{\rm{out}}}^2v_{{\rm{rel}}}^2}}{{r_s^2{{\left( {3{\rho _{{\rm{in}}}} + 2{\rho _{{\rm{out}}}}} \right)}^2}}} + \frac{{6{\rho _{{\rm{out}}}}{{\dot r}_s}{v_{{\rm{rel}}}}}}{{r_s^2\left( {3{\rho _{{\rm{in}}}} + 2{\rho _{{\rm{out}}}}} \right)}} - \frac{{6{\rho _{{\rm{in}}}} - 3{\rho _{{\rm{out}}}}}}{{3{\rho _{{\rm{in}}}} + 2{\rho _{{\rm{out}}}}}}\frac{{{a_{O'}}}}{{{r_s}}}\cos \vartheta 
	+ \frac{3}{4}\frac{{\dot r_s^2}}{{r_s^2}} - 10{\rm A}\frac{{{{\ddot r}_s}}}{{{r_s}}} + O{\left( \theta  \right)^2},
\end{equation}
where we made use of the asymptotic behaviour of the special functions $F_1$ and $F_2$ (see Appendix B). 
This result proves that to the linear-order $\hat \gamma^2$ is independent of the direction. This is an important theoretical result (not expected a priori) that indicates how our instability parameter is weakly dependent on the particular geometry developed (the curvature) and it has probably a wider range of applicability than what is formally mathematically permitted.

The importance of this result can be grasped by examining Fig. \ref{PiccoliAngoli}. If we assume that $\theta  \simeq 0$ is small (see point A on Fig. \ref{PiccoliAngoli}) then the difference between the orthogonal (and tangential) vectors to the unperturbed and perturbed surfaces is always small even if $\eta  \sim {r_s}$, i.e., even if $\eta  \ll {r_s}$  does not hold strictly (e.g., along the direction of the tidal tails of an orbiting dwarf galaxy with a highly radial orbit). Similarly, for the point B, i.e., far away from $\theta  = 0$  or $\theta  = \frac{\pi }{2}$  we expect the theory not to hold properly (as indeed the divergence of the special functions ${F_1}$  and ${F_2}$ indicates).
To grasp the importance of this result it is worth examining Fig. \ref{PiccoliAngoli}. If we assume that $\theta  \simeq 0$ is small (see point A on Fig. \ref{PiccoliAngoli}) then the difference between the orthogonal (and tangential) vectors to the unperturbed and perturbed surfaces is always small even if $\eta  \sim {r_s}$, i.e., even if $\eta  \ll {r_s}$  does not hold strictly (e.g., along the direction of the tidal tails of an orbiting dwarf galaxy with a highly radial orbit). Vice versa, for the point B, i.e., far away from $\theta  = 0$  or $\theta  = \frac{\pi }{2}$  we expect the theory not to hold properly (as indeed the divergence of the special functions ${F_1}$  and ${F_2}$ indicates).
If there were a strong angular dependence at the stagnation point to the first order (e.g., ${\theta ^2},{\theta ^3},{\theta ^4}...$) it would inevitably limit our results to the very specific spherical system (even though every stellar system with a sufficiently smooth density distribution can be approximated with an osculating sphere).

The ${\hat \gamma ^2}$'s independence in the first order in  $\theta $ is an encouraging result on the potential of the criterion we have developed. Finally, note that the angle $\vartheta $, differently from $\theta $,  is not necessarily small, but depends on the configuration space of the external cluster of galaxies or stellar distribution acting on the system under examination.

\subsection{Application to observations}
Although the investigation of a particular catalogue of galaxies, galaxy clusters, or globular cluster is beyond the goal of the present paper, it is interesting to observe that the criterion in Eq.\eqref{Eq37} can give a hint on the activity of the star formation directly by observational measures. Depending on the precision of the data available and on the system under study, the simplest approach to the criterion (from an observational point of view) is as follows. The mass spectrum of the molecular clouds, where stars are born, is reasonably well known (Paper I). For the external hot intergalactic medium the X-ray emitting hot intra-cluster gas distribution is known to be well represented by $\beta $-models \citep{1976A&A....49..137C}. Consider a galaxy with an effective radius ${r_s} = {r_{{\rm{eff}}}}$, starting to free fall in equilibrium ${\dot r_s} = {\dot r_{{\rm{eff}}}} = 0$ from the outskirts of a galaxy cluster parametrized by a potential-density couple $\Delta {\Phi _{{\rm{cl}}}} = 4\pi G{\rho _{{\rm{cl}}}}$. The galaxy will experience tidal stretching (or compression) by the tidal field ${\bm{T}} =  - \frac{{\partial {\Phi _{{\rm{cl}}}}}}{{\partial {\bm{x}}\partial {\bm{x}}}}$ along (or orthogonally) to the free fall direction (with velocity ${v_{{\rm{ff}}}} = v_{{\rm{rel}}}^ \bot $). Hence, it will be stable or unstable to star formation simply if the total mass ${M_{{\rm{gal}}}}$  of the galaxy is enough to shield the galaxy from the external field $a_{O'}^ \bot  = {\left. {\bm{T}} \right|_{{\rm{cl}}}}{r_s}$ or not (where ${\left. {\bm{T}} \right|_{{\rm{cl}}}}$ is evaluated through the radial direction to the galaxy cluster centre, ${\bm{a}} \simeq {{\bm{a}}_{O'}} + {{\bm{O}}^T}{\bm{TO\xi }} + ...$ ). The only observational datum required to be obtained is the tidal distribution of the gravitational system and this can easily be computed as 
\begin{equation}\label{Eq46bis}
	{{\rm T}_{ij}} = \sum\limits_{i \in {\rm{cl}}}^{} {\frac{{G{M_i}}}{{{{\left\| {{x_{{\rm{gal}}}} - {x_i}} \right\|}^3}}}\left( {\frac{{3\left( {{x_{{\rm{gal}}}} - {x_i}} \right)\left( {{x_{{\rm{gal}}}} - {x_j}} \right)}}{{{{\left\| {{x_{{\rm{gal}}}} - {x_i}} \right\|}^2}}} - {\delta _{ij}}} \right)},
\end{equation}
where $\bm{x}_\text{gal}$ is the location of the galaxy under examination within the catalogue describing the cluster of galaxies located respectively at $\bm{x}_i$, and $\delta _{ij}$ is the bi-dimensional Dirac delta function. An example of this type of computation from observational data is shown in \citet{2009A&A...499..385P} (for a different geometry than in Section \ref{SoR}). In this way, all the parameters necessary to exploit the instability criterion (e.g., in the form of Eq. \eqref{Eq44} ) are entirely obtained from a catalogue

\subsection{Numerical example on dwarf galaxies of the Local Group (LG)}
In paper I a local description of the instability processes was assumed, using a pressure equation (there Eq.(10)) that recovers standard literature results \citep[e.g.,][]{1972ApJ...176....1G} if considered in dimensionless systems (i.e., for ${r_s} \to 0$ we obtained the ram pressure equation of \citet{1972ApJ...176....1G}). This pressure equation was applied locally to a molecular cloud spectrum of masses $M \in \left[ {{{10}^2}{{,10}^6}} \right]{M_ \odot }$ \citep{1997ApJ...480..235E}. In this way, each different molecular cloud class was accounted differently for its instability (in the linear regime), reacting differently depending on the particular mass. The result was then integrated to obtain the overall mass consumed, transformed into stars, or transferred back to the intergalactic medium following the recipe of \citet{1999ApJ...516..619F}. The compatibility of the result was confirmed against a numerical integration of the evolution of an extensively studied LG dwarf galaxy (Carina) \citep{2011A&A...525A..99P}. 
In paper I a local description of the instability processes was assumed, using a pressure equation (there Eq.(10)) that recovers standard literature results \citep[e.g.,][]{1972ApJ...176....1G} if considered in dimensionless systems (i.e., for ${r_s} \to 0$ we obtained the ram pressure equation of \citet{1972ApJ...176....1G}). This pressure equation was applied locally to a molecular cloud spectrum of masses $M \in \left[ {{{10}^2}{{,10}^6}} \right]{M_ \odot }$ \citep{1997ApJ...480..235E}. In this way, each different molecular cloud class was accounted differently for its instability (in the linear regime), reacting differently depending on the particular mass. The result was then integrated to obtain the overall mass consumed, transformed into stars, or transferred back to the intergalactic medium following the recipe of \citet{1999ApJ...516..619F}. The compatibility of the result was confirmed against a numerical integration of the evolution of an extensively studied LG dwarf galaxy (Carina) \citep{2011A&A...525A..99P}. 

\begin{figure*}
\sidecaption
\includegraphics[width=\columnwidth]{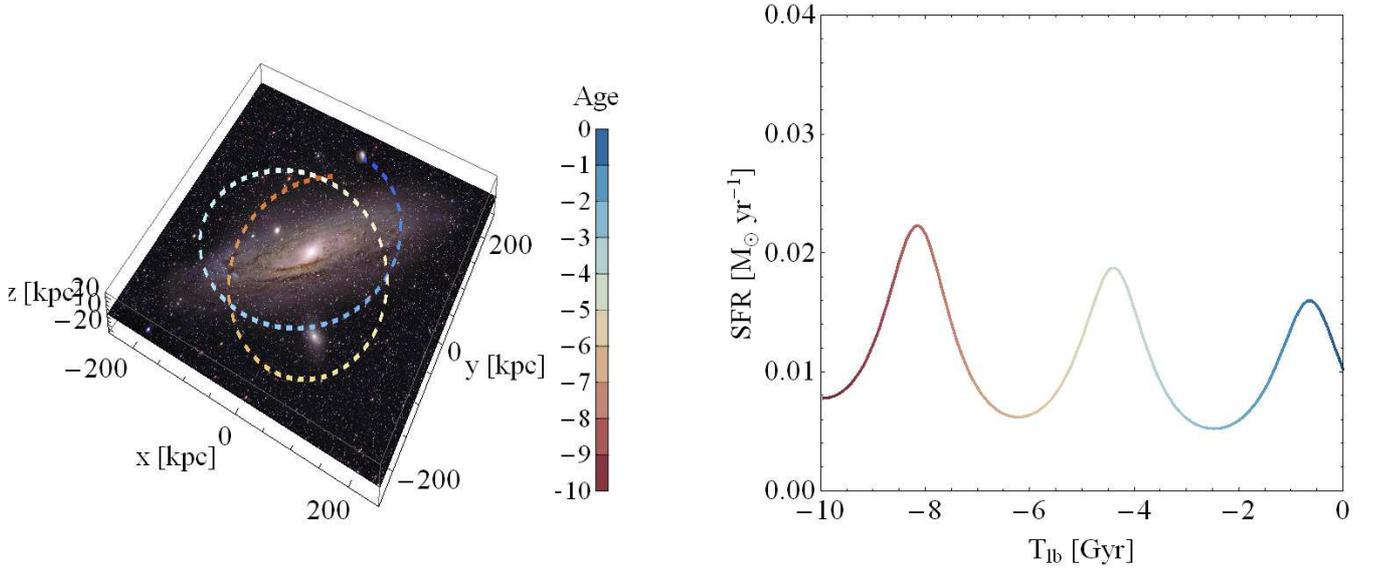}
\caption{(Left panel) Orbit of a dwarf galaxy with eccentricity e=0.25 and starting position $\left\{x,y,z\right\}=\left\{0,200,0\right\}$kpc computed on a MW tuned external potential (pictorial background photo). (Right panel) Star formation history of the dwarf galaxy of the left panel accounting for internal \textit{and }external effects as modelled in Paper I.}  
\label{OrbSFHe25}
\end{figure*}

With the criterion of instability derived above in Eq. \eqref{Eq36_2}, we can now investigate more precisely the role of the different orbital parameters involved in the instability process. For example, we assume a dwarf galaxy orbiting in the plane of the MW potential, starting at 200 kpc from the centre of ${S_0}$ (centred on the MW) on an orbit with eccentricity $e = 0.25$. The orbit and star formation (for an initially metal poor galaxy) is as in Fig.\ref{OrbSFHe25}
where in the left panel the orbit computed for the MW galaxy model of Paper I is illustrated. The legend explains the colour-coding of the orbits as a function of time (Gyrs). The same colour-code is used in the right panel where the star formation history has been computed with the technique developed in Paper I. 

\begin{figure*}
\sidecaption
\includegraphics[width=\columnwidth]{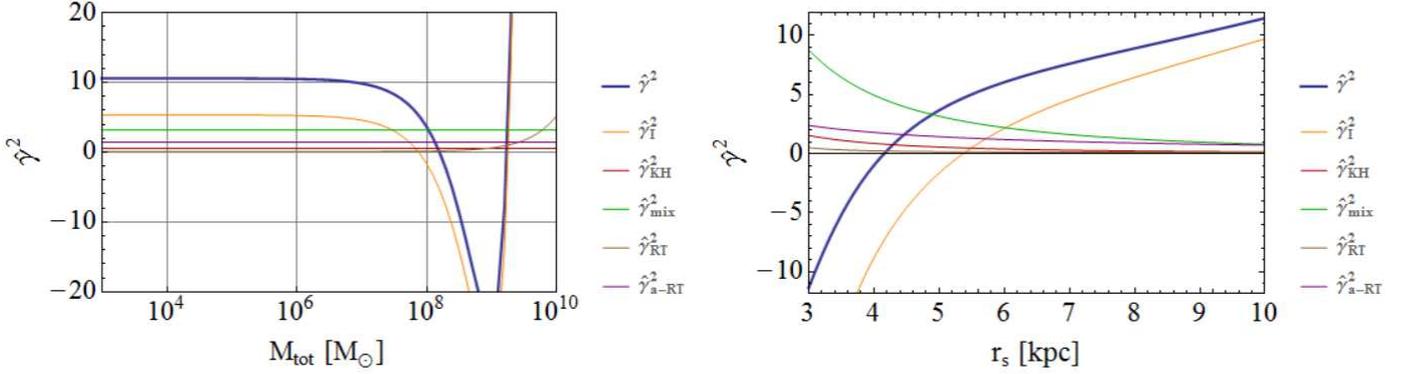}
\caption{(Left panel) The instability parameter as a function of the mass of the dwarf galaxy orbiting in a MW external environmental model (see text for details). (Right panel) Instability parameter as a function of the size for a stable mass chosen the left panel, $M = {10^8}{M_ \odot }$. The different thin lines and colours refer to the different components contributing to the global instability parameter (thick blue line).}  
\label{GraphGrid}
\end{figure*}

For an example position along the orbit, say ${t^*} \equiv t_{\text{lbt}} = -9$ Gyr (where $t_{\text{lbt}}$ is the look-back time), we ask ourselves which mass limit gives rise to star formation instability. We plot our instability factor Eq.\eqref{Gamma2parameter}, with the model of the MW external potential and electron number density for coronal gas as in Paper I, as a function of the total mass of the dwarf galaxy. The results are shown in Fig.\ref{GraphGrid} (left panel).
The orbits define the phase space parameter of the galaxy. If we increase the total mass of the orbiting object at fixed orbital parameters (MW model centred on $S_0$, ${\mathbf{x}}\left( {t = 0} \right) = \left\{ {0,200,0} \right\}$kpc and $e = 0.25$), we see that the system gradually becomes more stable and at $t=t^*$ (an arbitrary point on the real line of the time) we can easily see that systems  more massive than $\sim 0.25 \times {10^8}{M_ \odot }$ with a tidal radius of ${r_{{\text{tidal}}}}=5$kpc become stable to external star formation activation.

We remark at this point that Eq.\eqref{Eq37} represents a criterion of stability, not an equation governing the evolution of the system analysed. We do not follow the linear response of a system but only study the onset of star formation. Different masses, or systems, would evolve on different orbits than the one chosen in our example. The instability criterion simply has the function to predict which parameters in the multidimensional space of mass, size and phase-space give rise to instability and hence potentially lead to star formation because of the specified external environment. Of course if the galaxy does not contain gas (i.e., the criterion of $\Xi>0$ in Eq.\eqref{Eqsys01} is not satisfied), then regardless of whether its structural and orbital parameters satisfy the positivity of the instability growth factor, the galaxy will not experience star formation. More on this point will be said in the Section 5.

Another interesting feature of our theory is the possibility to account for a scale parameter ${r_s}$. Hence, on the same fixed orbit of Fig.\ref{GraphGrid} (left panel) we can investigate the instability once ${r_s}$ is allowed to vary. We consider the same instant and orbit. As seen in the left panel of Fig.\ref{GraphGrid}, any mass ${M_{{\rm{gal}}}} > {0.25 \times 10^8}{M_ \odot }$ is enough to shield the galaxy from activating star formation by external factors. We now imagine to dilute (or compact) a ${M_{{\rm{gal}}}} = {10^8}{M_ \odot }$ over larger and larger scale radii at the same position and velocity as computed for ${t^*}$ in the previous plot. The result is as in Fig.\ref{GraphGrid} (right panel). The result clearly shows that the growth of the instabilities is favoured by diffusing the stellar system. As soon as a galaxy of total mass ${M_{{\rm{gal}}}} = {10^8}{M_ \odot }$ is diffused over a scale radius greater than 4.1 kpc the galaxy becomes prone to the growth of instabilities (here the tidal radius, but note that the passage between different scale radii will result in just a shift along the x axis) coherently with left panel in the figure.

Finally, both the panels show a comparative study in the particular instant of the selected orbit for the relative importance of the different effects. We selected an angular dependence of $\theta  = \frac{\pi }{8}$ to show all the effects contributing to model the total instability parameter curve. As is evident, the mixed term $\hat \gamma _{{\rm{mix}}}^2$ is dominant over the pure KH term, $\hat \gamma _{KH}^2$, and RT term, $\hat \gamma _{{\rm{RT}}}^2$. This holds for compact systems. At fixed mass ($M = {10^6}{M_ \odot }$) for increasing radius, we see that is the more diffuse is the system, the more the inertial term of Eq.\eqref{Eq38} becomes relevant. It finally becomes dominant over 5 kpc. We stress once more that this is not expected to be a general trend, but it is specific to this particular orbit. Nevertheless, for each orbit, the instability criterion can indicate the dominant effects for the parameter selected. The RT-acceleration effect (Eq.\eqref{Eq42}) is constant at a fixed point on the orbit and dominant over all terms. This is because for the chosen orbits and dwarf scale parameters $r_s=5$ kpc and $M_\text{gal}=10^8 M_\odot$ there is a tight correlation between pericentre passages and star formation history (see Fig. \ref{OrbSFHe25} right panel).

\begin{figure*}
\sidecaption
\includegraphics[width=12cm]{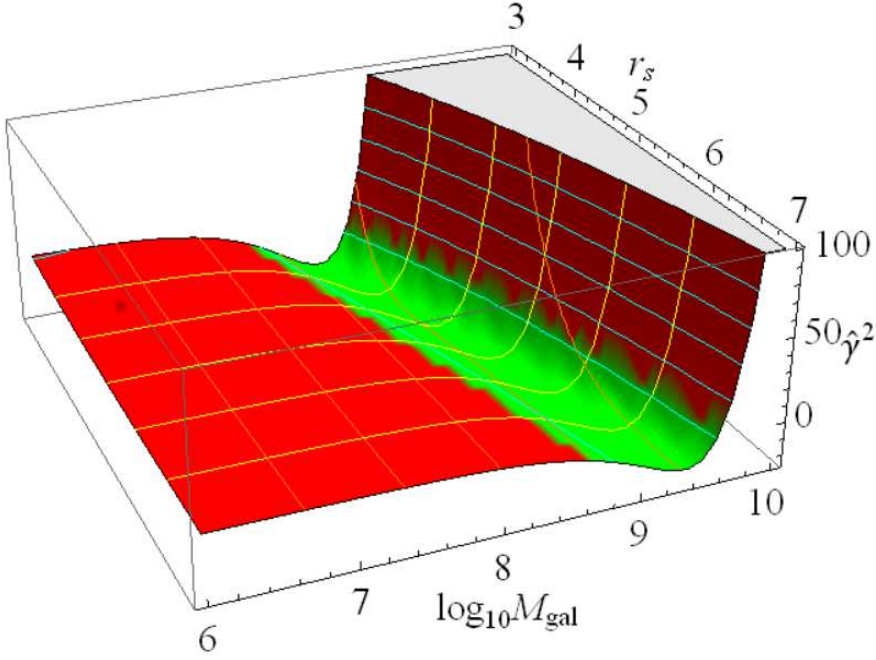}
\caption{Instability valley: manifold of the star formation instability for a ${\mathbf{x}}\left( {t = 0} \right) = \left\{ {0,200,0} \right\}$kpc and $e = 0.25$ considered as an example. The green zone refers to stable regions of the mass vs. size space. Red zones refer to possible active star formation.}
\label{GammaTot}
\end{figure*}

We combine the two panels of the previous figures to show in Fig.\ref{GammaTot} (a given orbit and precise instant, $t=t^*$ in our case) the characteristic manifold of the star forming regions (in the mass-size space).
As evident from the Figure, the dwarf galaxy orbiting the MW in the example orbit can either have star formation (red zone) or be stable against it (green region) depending on its mass and size. \textit{The criterion derived here predicts the threshold value for the onset of star formation in a mass vs. size space for any orbit of interest. }This plot indeed can be calculated for to any point of the MW dwarf galaxies' phase-space distribution. In Fig. \ref{GammaTot} the ``green valley'' of the manifold formalizes the intuition that smaller (in size) systems require less total mass to be shielded from external influences. Finally, beyond a certain limit the internal-inertial term $\hat \gamma^2_I$ becomes dominant and induces the $l=2$ modal instability regardless of the role of the external pressure or tidal forces. 
As evident from the Figure, the dwarf galaxy orbiting the MW in the example orbit can either have star formation (red zone) or be stable against it (green region) depending on its mass and size. \textit{The criterion derived here predicts the threshold value for the onset of star formation in a mass vs. size space for any orbit of interest. }This plot indeed can be calculated for to any point of the MW dwarf galaxies' phase-space distribution. In Fig. \ref{GammaTot} the ``green valley'' of the manifold formalizes the intuition that smaller (in size) systems require less total mass to be shielded from external influences. Finally, beyond a certain limit the internal-inertial term $\hat \gamma^2_I$ becomes dominant and induces the $l=2$ modal instability regardless of the role of the external pressure or tidal forces.

We stress that this is not intended to be an investigation of the instability zones of the parameter spaces of the MW, LG or any particular LG dwarf galaxy. A statistical investigation of the errors involved and on their propagation on the positions and  velocities of a dwarf galaxy is a complicated task that requires more advanced techniques (e.g., see the analysis of the Carina’s dwarf galaxy orbit in \citet{2011A&A...525A..99P} based on the minimum action principle) and is in preparation for MW dwarf galaxies (Pasetto et al 2014, in preparation).

\section{Conclusion}
Since the original works on collapse and instability of Jeans \citep[][]{1902RSPTA.199....1J} and the phenomenological works of \citet[][]{1959ApJ...129..243S}, and \citet[][]{1998ApJ...498..541K}, criteria ruling the star formation processes have been of great interest in astrophysics and an extensive matter of debate. The treatment of the star formation processes accounting for environmental effects has almost always been the “territory” of experimental/numerical astrophysics (for a review, see e.g., \citet[][]{2010AdAst2010E..25M}).

In this work we address this problem from an analytical point of view for the first time, by presenting a new treatment of the gas instability processes that activate star formation in interacting stellar systems embedded in an external environment. 
Our approach is based on the study of the pressure acting on a density distribution of molecular clouds subject to external pressure acting on them. The arguments are developed in spherical geometry and a consistent new instability criterion is obtained, which accounts for gravitational and hydrodynamical properties of the molecular clouds and their surrounding environment. 

The main results in this analysis are 

\begin{itemize}
	\item an analytical expression for the instability conditions, a criterion obtained by analysing the growth of the instability because of a perturbation at a surface of equilibrium. The description of the perturbation is limited to a mode of interest for the astrophysical case, $l = 2$ in a spherical harmonic expansion. Limiting conditions (Eqs.\eqref{Eq43} or \eqref{Eq44} with \eqref{Eq48}) are also developed to propose a version of the instability criterion fully testable with limited observational data. \textit{ From observational constraints this is expected to give indications on the nature of a studied system}.
	\item we show for the first time the dependence of instability on the orbital parameters for a particular example. This approach has the advantage of casting light on the role of the different instability processes in giving rise to unstable (i.e. growth) modes. In particular, this approach is complementary to pure numerical methodology (adaptive mesh refinement, smooth particle hydrodynamics etc.) indicating the main dependencies of the analysed stellar system on dynamical parameters (speed, acceleration, mass and size) and how strong they are. In this way, this analytical result is a fundamental key for interpreting the numerical/experimental results where all of these effects act simultaneously.
\end{itemize}

Finally, an investigation of the star formation criterion is presented for a typical orbit of a LG dwarf galaxy \citep[e.g.,][]{2014ApJ...789..147W, 2009ARA&A..47..371T, 1997RvMA...10...29G}. We briefly presented our instability parameter as an investigative tool to the stellar formation in pre-assigned orbits of a LG dwarf galaxy.

We conclude with a few remarks on the criterion developed here. 
It is obtained by a dynamically consistent equation derived to the first order in the perturbation factor $\eta $, but it is not an evolution equation, it is only an instability criterion. We did not consider the eigen-function of the perturbation equation nor the equilibrium equation that should be solved together to obtain the time evolution of the perturbation, eventually producing a system of equations directly comparable to N-body AMD/SPH simulations. This comparison was done in Paper I to which we refer the reader.

Despite the difficulties in handling instability with numerical techniques, the N-body AMD/SPH simulations provide a valuable tool to perform experiments/exercises that can guide the theoretical and observational studies. Nevertheless, the degree of comprehension of a phenomenon that can be achieved with analytical studies cannot be reached by a controlled numerical experiment where all the effects (internal and external) overlap in a non-linear way. It is the purpose of this paper to present a possible interpretative key for disentangling the different theoretical aspects of a numerical experiment.

Still, numerical experiments can lead the theoretical research where the limitation of the analytical approaches struggles to advance to a simple formulation \citep[e.g.,][]{2012MNRAS.419..971D}. For example every time two systems lose their identities when merging into a single object \citep[e.g.,][]{2014MNRAS.442L..33R, 2008MNRAS.391L..98R,2007MNRAS.379.1475S} the linear response theory here developed can be only of indicative help, while a numerical simulation where the instability criteria are implemented locally seems - to date - the better way to do advance our understanding.

This work addresses in an analytical way the problem of the environmental influence on a system. The criterion derived here predicts the threshold value for the onset of star formation in a mass vs. size space for any orbit of interest. It shows that the instability can be triggered or suppressed in a different way depending on the internal density profile of the system under examination. 
Finally, we remark that in the case of primordial globular clusters moving supersonically throughout the disk of a spiral galaxy, the algebra of the instability criterion developed here is expected to work once the pressure equation is considered in the supersonic regime. A relation between pre and post shock pressure to account for this isentropic compression was already worked out in Appendix A.3 of Paper I.



\begin{acknowledgements}
SP thanks: Anna Pasquali and Denija Crnojevic for stimulating discussions. We thank the anonymous referee for useful suggestions. We acknowledge PRACE for awarding us access to resource ARCHER UK National Supercomputing Service. The support of Chris Johnson from EPCC, UK for the technical work is gratefully acknowledged.
 
\end{acknowledgements}


\appendix
\section{Non-inertial linear response theory for gas instabilities in spherical coordinates}\label{Linearresponsetheory}
In this appendix we develop the mathematics of the linear response theory introduced in Section \ref{LRT}. The unstable periphery of a gaseous sphere penetrating an external medium is considered in a non-inertial reference frame and kinematic and dynamical boundary conditions are considered. An instability criterion is obtained in spherical coordinates and plane geometry limit is considered.

\subsection{Kinematic boundary conditions}\label{Kinematicboundaryconditions}

\subsubsection{Perturbed surface}\label{Perturbedsurface}
In the geometrical framework introduced in the Section \ref{GeometricalSoR}, we consider a potential-flow type description of the surface of the galaxy in its motion throughout an intra-cluster medium The distribution of the molecular clouds in the interstellar medium of the galaxy is described with a density distribution $\rho  = \rho \left( {\bm{\xi }} \right)$ in ${S_1}$ bordered by a surface $\Sigma$ surface of the frontier of the domain of existence of the (bound) density function $\rho :\mathop {\lim }\limits_{{\bm{\xi }} \to \infty } \rho  < \infty $. No singularity is allowed in the potential-density couple satisfying the associated Poisson equation $\Delta \Phi  = 4\pi G\rho $. This distribution is then perturbed to a new state, corresponding to a new perturbed surface density (where the spherical coordinates introduced above in ${S_1}$ have been employed). We will limit ourselves to a linear analysis and we assume the defining equation for the surface $\Sigma \left( {\xi ,\theta ,\phi ;t} \right)=0 $ to be given by Eq.\eqref{Eq02}.

\subsubsection{Internal gas perturbed potential flow}\label{Internalgasperturbedpotentialflow}
We will refer to a quantity of the orbiting stellar system as ``internal'', e.g., its density ${\rho _{{\rm{in}}}}$, velocity potential ${\varphi ^{{\rm{in}}}}$ etc. To describe the cold interstellar medium we will use the solution for the Laplace equation for a stationary expanding/contracting potential flow written as ${\varphi _{{{\bm{v}}_1}}^{III}} \equiv  - \frac{{r_s^2{{\dot r}_s}}}{\xi }$ \citep[e.g.,][]{ 1959flme.book.....L} to which we add the perturbation solution of the Laplace equation proportional to ${\xi ^l}$, i.e. ${\xi ^l}Y_l^m{B_{lm}}$ (with $B_{lm}$ proportionality coefficients of the spherical harmonic basis):
\begin{equation}\label{Eq03}
	\varphi _{{{\bm{v}}_1}}^{{\text{in}}}\left( {\xi ,\theta ,\phi ;t} \right) = {\xi ^l}Y_l^m{B_{lm}} - \frac{{r_s^2{{\dot r}_s}}}{\xi },
\end{equation}
where we already excluded terms proportional to ${\xi ^{ - 1 - l}}$ in the radial solution of the Laplace equation $\varphi  \propto {A_{lm}}{\xi ^{ - l}} + {B_{lm}}{\xi ^l}$ by setting their corresponding coefficients ${A_{lm}} = 0$. This is done in order to avoid divergences as long as we move away from $\Sigma $ inward into the galaxy. To ensure continuity of the surface element fluids at the surface, we proceed in the standard way \citep[e.g.,][]{2000ifd..book.....B} by evaluating the kinematical boundary conditions (i.e., of the Eulerian derivative at the surface) of the fluid elements at the perturbed surface $\left\| {\bm{\xi }} \right\| = {r_s} + \eta Y_l^m$ (see Eq.\eqref{Eq02}):
\begin{equation}\label{Eq04}
	{\left. {{\partial _t}\Sigma  + \left\langle {\nabla \varphi _{{{\bm{v}}_1}}^{{\text{in}}},\nabla \Sigma } \right\rangle } \right|_{\xi = \eta Y_l^m + {r_s}}} = 0,
\end{equation}
where ${\partial _x}$ is a compact notation for the derivative $\frac{\partial }{{\partial x}}$. With ${\partial _t}\Sigma  =  - {\dot r_s} - \dot \eta Y_l^m$ and by computing the spatial gradient components in $S_1$ as $\left( {{\partial _\xi }\Sigma ,\frac{{{\partial _\theta }\Sigma }}{\xi },\frac{{{\partial _\phi }\Sigma }}{{\xi \sin \theta }}} \right) = \left( {1, - \frac{{\eta {\partial _\theta }Y_l^m}}{\xi }, - \frac{{\eta {\partial _\phi }Y_l^m}}{\xi }\csc \theta } \right)$, as well as $\left( {{\partial _\xi }\varphi _{{{\bm{v}}_1}}^{{\text{in}}},\frac{{{\partial _\theta }\varphi _{{{\bm{v}}_1}}^{{\text{in}}}}}{\xi },\frac{{{\partial _\phi }\varphi _{{{\bm{v}}_1}}^{{\text{in}}}}}{{\xi \sin \theta }}} \right) = \left( {l{\xi ^{l - 1}}Y_l^m{B_{lm}} + \frac{{r_s^2{{\dot r}_s}}}{{{\xi ^2}}},{B_{lm}}{\xi ^{l - 1}}{\partial _\theta }Y_l^m,{B_{lm}}{\xi ^{l - 1}}\csc \theta {\partial _\phi }Y_l^m} \right)$, Eq.\eqref{Eq04} reduces to an equation for the parameters ${B_{lm}}$:
\[{B_{lm}}\eta {\left( {{\partial _\theta }Y_l^m} \right)^2}{\left( {\eta Y_l^m + {r_s}} \right)^{l - 2}} + {B_{lm}}\eta {\csc ^2}\theta {\left( {{\partial _\phi }Y_l^m} \right)^2}{\left( {\eta Y_l^m + {r_s}} \right)^{l - 2}} - {B_{lm}}lY_l^m{\left( {\eta Y_l^m + {r_s}} \right)^{l - 1}} = \frac{{r_s^2{{\dot r}_s}}}{{{{\left( {\eta Y_l^m + {r_s}} \right)}^2}}} - \dot \eta Y_l^m - {\dot r_s},\]
 obtained by Eq.\eqref{Eq04} with the terms computed above and by simple substitution of the perturbed surface of Eq.\eqref{Eq02}.  This equation can be solved for ${B_{lm}}$ as:
\begin{equation}\label{Eq05}
{B_{lm}} = \frac{{Y_l^m{{\left( {\eta Y_l^m + {r_s}} \right)}^{ - l}}\left( {\dot \eta {{\left( {\eta Y_l^m + {r_s}} \right)}^2} + \eta {{\dot r}_s}\left( {\eta Y_l^m + 2{r_s}} \right)} \right)}}{{ - \eta \left( {{{\left( {{\partial _\theta }Y_l^m} \right)}^2} + {{\csc }^2}\theta {{\left( {{\partial _\phi }Y_l^m} \right)}^2}} \right) + l{r_s}Y_l^m + \eta l(Y_l^{m})^2}},
\end{equation}
obtained by collecting the common terms. We now linearize the previous result to the first order in $\eta $. After a McLaurin expansion in $\eta$ we find the following compact form for the coefficients ${B_{lm}}$:
\begin{align}\label{Eq06}
{B_{lm}}\left( \eta  \right) &\simeq {B_{lm}}\left( 0 \right) + {\partial _\eta }{B_{lm}}\left( 0 \right)\eta  + O{\left( \eta  \right)^2} ,\nonumber\\
&\simeq \dot \eta \frac{{r_s^{1 - l}}}{l} + 2\eta {{\dot r}_s}\frac{{r_s^{ - l}}}{l}. 
\end{align}
Inserting Eq.\eqref{Eq06} in Eq.\eqref{Eq03} helps us to obtain the final form of the potential vector to the first order as:
\begin{equation}\label{Eq07}
	\varphi _{{{\bm{v}}_1}}^{{\text{in}}}\left( {\xi ,\theta ,\phi ;t} \right) \simeq {\xi ^l}Y_l^m\left( {\frac{{\dot \eta }}{l}{r_s}^{1 - l} + \frac{{2\eta }}{l}{{\dot r}_s}{r_s}^l} \right) - \frac{{r_s^2{{\dot r}_s}}}{\xi },
\end{equation}
which is Eq.(4) of \citet[][]{1954JAP....25...96P}. Differently from \citet[][]{1954JAP....25...96P}, we are here interested in describing the motion of the dwarf galaxy along its orbit in the bath of a hotter, lighter intergalactic medium, or vice versa, the motion of this intergalactic medium impacting the dwarf galaxy in its orbital evolution as it appears in the reference frame ${S_1}$. This case has similarity with the problem recently presented in Paper I and was there extensively treated in the context of stellar convection by \citet[][]{2014arXiv1403.6122P}. We adapt their formalism and extend their results to this non-axisymmetric context.

\subsubsection{External gas perturbed potential flow}\label{Externalgasperturbedpotentialflow}
We will refer to a quantity external to the orbiting system as ``outside'' the system, e.g., the hot intra-cluster medium density ${\rho _{{\rm{out}}}}$, its velocity potential ${\varphi ^{{\rm{out}}}}$ etc. The potential flow for the hot intergalactic medium written in the reference frame ${S_1}$ comoving with the stellar system, $\varphi _{{{\bm{v}}_1}}^{{\text{out}}}$, is introduced in the previous section, but see recently also \citet[][]{2014arXiv1403.6122P}, as ${\varphi _{{{\bm{v}}_1}}} \equiv  - v\xi \left( {1 + \frac{1}{2}\frac{{r_s^3}}{{{\xi ^3}}}} \right)\cos \theta $. To these terms, we add now the term computed above for the expansion/contraction of the galaxy ${\varphi _{{{\bm{v}}_1}}}$, and the perturbation solution of the Laplace equation proportional to $\frac{{1}}{{{\xi ^{l + 1}}}}$  written as ${A_{lm}}\frac{{Y_l^m}}{{{\xi ^{l + 1}}}}$ to get
 \begin{equation}\label{Eq08}
	 \varphi _{{{\bm{v}}_1}}^{{\text{out}}}\left( {\xi ,\theta ,\phi ;t} \right) =  - v\xi \left( {1 + \frac{1}{2}\frac{{r_s^3}}{{{\xi ^3}}}} \right)\cos \theta  - \frac{{r_s^2{{\dot r}_s}}}{\xi } + {A_{lm}}\frac{{Y_l^m}}{{{\xi ^{l + 1}}}},
 \end{equation}
where differently from the previous case of Eq.\eqref{Eq03} we want here to exclude terms proportional to ${\xi ^l}$ by setting their corresponding coefficients ${B_{lm}} = 0$ in the Laplace equation because we do not want to consider divergences as long as we go far outside the dwarf galaxy away from $\Sigma $. Again as in Eq.\eqref{Eq04} we proceed by evaluating the kinematical boundary conditions of the fluid element at the surface 
\begin{equation}\label{Eq09}
	{\left. {{\partial _t}\Sigma  + \left\langle {\nabla \varphi _{{{\bm{v}}_1}}^{{\text{out}}},\nabla \Sigma } \right\rangle } \right|_{{\xi} = \eta Y_l^m + {r_s}}} = 0,
\end{equation}
where the only difference from the previous Eq.\eqref{Eq04} is that the velocity potential gradients are now derived as $\left( {{\partial _\xi }\varphi _{{{\bm{v}}_1}}^{{\text{out}}},\frac{1}{\xi }{\partial _\theta }\varphi _{{{\bm{v}}_1}}^{{\text{out}}},\frac{{{\partial _\phi }\varphi _{{{\bm{v}}_1}}^{{\text{out}}}}}{{\xi \sin \theta }}} \right)$ $ = \left( {\frac{{3r_s^3}}{{2{\xi ^3}}}v\cos \theta  - \left( {1 + \frac{{r_s^3}}{{2{\xi ^3}}}} \right)v\cos \theta  - {A_{lm}}\left( {l + 1} \right){\xi ^{ - l - 2}}Y_l^m + \frac{{r_s^2{{\dot r}_s}}}{{{\xi ^2}}},} \right.$  $v\left( {\frac{{r_s^3}}{{2{\xi ^3}}} + 1} \right)\sin \theta  + {A_{lm}}{\xi ^{ - l - 2}}{\partial _\theta }Y_l^m,$ $\left. {{A_{lm}}{\xi ^{ - l - 2}}\csc \theta {\partial _\phi }Y_l^m} \right)$. Considering this difference, we proceed exactly as done above for the $\varphi _{{{\bm{v}}_1}}^{{\text{in}}}$, to obtain an equation that is linear in ${A_{lm}}$ and that can be solved as:
\begin{equation}\label{Eq10}
{A_{lm}} =  - \frac{{{{\left( {\eta Y_l^m + {r_s}} \right)}^{l + 3}}\left( {\frac{{\eta v{\partial _\theta }Y_l^m\left( {2{{\left( {\eta Y_l^m + {r_s}} \right)}^3} + r_s^3} \right)\sin \theta }}{{2{{\left( {\eta Y_l^m + {r_s}} \right)}^4}}} + v\left( {1 - \frac{{r_s^3}}{{{{\left( {\eta Y_l^m + {r_s}} \right)}^3}}}} \right)\cos \theta  + {{\dot r}_s}\left( {1 - \frac{{r_s^2}}{{{{\left( {\eta Y_l^m + {r_s}} \right)}^2}}}} \right) + \dot \eta Y_l^m} \right)}}{{\eta \left( {{{\left( {{\partial _\theta }Y_l^m} \right)}^2} + {{\csc }^2}\theta {{\left( {{\partial _\phi }Y_l^m} \right)}^2} + (l + 1){{\left( {Y_l^m} \right)}^2}} \right) + (l + 1){r_s}Y_l^m}}.
\end{equation}
We now linearize the previous result to the first order in $\eta $. After some algebra we get:
\begin{equation}\label{Eq11}
		{A_{lm}} \simeq  - 3\eta v\frac{{r_s^{l + 1}{\partial _\theta }Y_l^m}}{{2(l + 1)Y_l^m}}\sin \theta  - 2\eta {\dot r_s}\frac{{r_s^{l + 1}}}{{l + 1}} - \dot \eta \frac{{r_s^{l + 2}}}{{l + 1}} - 3\eta v\frac{{r_s^{l + 1}}}{{l + 1}}\cos \theta . 
\end{equation}
Finally, we obtain the potential velocity in the following simplified form:
\begin{equation}\label{Eq12}
\varphi _{{{\bm{v}}_1}}^{{\text{out}}}\left( {\xi ,\theta ,\phi ;t} \right) \simeq  - v\cos \theta \xi \left( {1 + \frac{1}{2}\frac{{r_s^3}}{{{\xi ^3}}}} \right) - \frac{{r_s^2{{\dot r}_s}}}{\xi } - \frac{{3{r_s}^{l + 1}}}{{\left( {l + 1} \right){\xi ^{l + 1}}}}\eta \left( {\frac{{v{\partial _\theta }Y_l^m}}{2}\sin \theta } \right.\left. { + Y_l^m\left( {\frac{1}{3}\frac{{\dot \eta }}{\eta }{r_s} + \frac{2}{3}{{\dot r}_s} + v\cos \theta } \right)} \right),
\end{equation}
where we have inserted Eq.\eqref{Eq11} in Eq.\eqref{Eq08} and accepted minor simplifications.

A sanity check shows that this can be reduced in the unperturbed case, $\eta  \to 0$, to the result ${\varphi _{{{\bm{v}}_1}}} + {\varphi _{{{\bm{v}}_1}}^{III}}$ This was already suggested in Paper I and extensively considered in a different context in \citet[][]{2014arXiv1403.6122P}. For $v = 0 \wedge \eta  \ne 0$ this reduces to the Eq. (5) of \citet[][]{1954JAP....25...96P}.

\subsection{Dynamical boundary condition}\label{Dynamicalboundarycondition}
At the surface radius $\left\| {\bm{\xi }} \right\| = {r_s} + \eta Y_l^m$ that we have chosen to represent the galaxy size, apart from the kinematic boundary condition we want to express the condition of continuity of the stress vector (i.e. the dynamic boundary condition). The stress vector ${{\bm{s}}_{{\text{out}}}}$ and ${{\bm{s}}_{{\text{in}}}}$ inside and outside the surface $\Sigma$ must satisfy the condition ${\left\langle {{\bm{n}},{{\bm{s}}_{{\text{out}}}}} \right\rangle _{\Sigma  = 0}} = {\left\langle {{\bm{n}},{{\bm{s}}_{{\text{in}}}}} \right\rangle _{\Sigma  = 0}}$  so that for inviscid fluids (${\bm{s}} =  - p{\bm{I}}$ with ${\bm{I}}$ identity matrix) we obtain the standard literature dynamical boundary condition ${p_{{\text{out}}}} = {p_{{\text{in}}}}$ to be treated now thus accounting for the ram pressure that the galaxy is experiencing in its motion.  

\subsubsection{Internal gas pressure equation}\label{Internalgaspressureequation}
Now we need the task to impose the dynamical boundary condition of the external gas medium on the internal stellar system gas at each position of its perturbed scale-radius surface ${r_s} + \delta {r_s}$. Given the framework developed in Section \ref{GeometricalSoR} we can make use of Eq.(7) of Paper I where the non-inertial character of the system ${S_1}$ is taken into account. In this notation, we can compute the velocity for the molecular clouds of the dwarf galaxy $\left\| {{{\bm{v}}_1}} \right\|$ as ${\left\| {{{\bm{v}}_1}} \right\|^2} = \left\langle {{\nabla _{\bm{\xi }}}\varphi _{{{\bm{v}}_1}}^{{\text{in}}},{\nabla _{\bm{\xi }}}\varphi _{{{\bm{v}}_1}}^{{\text{in}}}} \right\rangle $ is  the internal fluid of the galaxy which is inert with respect to the reference frame ${S_1}$ comoving with the stellar system. Eq.(7) of Paper I in this case reads:
\begin{equation}\label{Eq13}
{\partial _t}\varphi _{{{\bm{v}}_1}}^{{\text{in}}} + \frac{1}{2}\left\langle {\nabla \varphi _{{{\bm{v}}_1}}^{{\text{in}}},\nabla \varphi _{{{\bm{v}}_1}}^{{\text{in}}}} \right\rangle  + \frac{p}{{{\rho _{{\text{in}}}}}} = {f^{{\text{in}}}}\left( t \right) - {\Phi _{\bm{g}}} - \left\langle {{{\bm{a}}_{O'}},{\bm{\xi }}} \right\rangle ,
\end{equation}
where $\left\langle {{{\bm{a}}_{O'}},{\bm{\xi }}} \right\rangle$ is the projection of the acceleration along the position vector ${\bm{\xi }}$ and ${f^{{\text{in}}}}\left( t \right)$ is a constant of the space, not depending on ${B_{{\text{lm}}}}$, which we determine by imposing the boundary condition far away from the ideal radius ${r_s}$ (at infinity). This is because we assume hydrostatic equilibrium far away from the molecular cloud borders and the function ${f^{{\text{in}}}}$ is therefore determined by the limit of the previous equation for $\left\| {\bm{\xi }} \right\| \to \infty$ as shown in Paper I.

Because the lifetime of a molecular cloud (given the star formation efficiency expected to act in the systems under study, see Fig. 1 of Paper I) is much shorter ($<300$ Myr) than the timescale over which the orbital parameters change significantly, we assume the velocity of the fluid impacting the dwarf galaxy molecular clouds to be uniform and constant (in ${S_1}$). We also neglect non-orthogonal components of the acceleration that remain constant in time along the lifetime of the molecular clouds. In this case we write simply $\left\langle {{{\bm{a}}_{O'}},{\bm{\xi }}} \right\rangle  = {a_{O'}}\xi \cos \vartheta $ with $\vartheta $ being the angle between ${{\bm{a}}_{O'}}$ and ${\bm{\xi }}$. In general $\vartheta  \ne \theta $ apart from particular orbits (or part of them).

To make progress with Eq.\eqref{Eq13} we need to evaluate ${\left. {{\partial _t}\varphi _{{{\bm{v}}_1}}^{{\text{in}}}} \right|_{\left\| {\bm{\xi }} \right\| = \eta Y_l^m + {r_s}}}$ and ${\left. {\left\langle {\nabla \varphi _{{{\bm{v}}_1}}^{{\text{in}}},\nabla \varphi _{{{\bm{v}}_1}}^{{\text{in}}}} \right\rangle } \right|_{\left\| {\bm{\xi }} \right\| = \eta Y_l^m + {r_s}}}$  to first order in the small parameter $\eta $. Differentiating Eq. \ref{Eq07} gives 
\begin{equation}\label{Eq14}
	{\partial _t}\varphi _{{{\bm{v}}_1}}^{{\text{in}}} = \frac{{{\xi ^l}Y_l^mr_s^{ - l}}}{l}\left( {\ddot \eta {r_s} + 2\eta {{\ddot r}_s} + 3\dot \eta {{\dot r}_s}} \right) - {\xi ^l}{\dot r_s}Y_l^mr_s^{ - l - 1}\left( {\dot \eta {r_s} + 2\eta {{\dot r}_s}} \right) - \frac{{r_s^2{{\ddot r}_s}}}{\xi } - \frac{{2\dot r_s^2{r_s}}}{\xi },
\end{equation}
to be evaluated at the perturbed location $\xi  = \eta Y_l^m + {r_s}$. We expand this to the first order to obtain: 
\begin{align}\label{Eq15}
  {\left. {{\partial _t}\varphi _{{{\bm{v}}_1}}^{{\text{in}}}} \right|_{\xi = \eta Y_l^m + {r_s}}} &\simeq \ddot \eta {r_s}\frac{{Y_l^m}}{l} + 3\dot \eta {{\dot r}_s}\frac{{Y_l^m}}{l} - \dot \eta {{\dot r}_s}Y_l^m - {r_s}{{\ddot r}_s} - 2\dot r_s^2 \hfill \nonumber\\
   &+ \eta \left( {\frac{{3\dot \eta {{\dot r}_s}}}{{{r_s}}}{{\left( {Y_l^m} \right)}^2} - \frac{{\dot \eta {{\dot r}_s}}}{{{r_s}}}l{{\left( {Y_l^m} \right)}^2} + \frac{{2Y_l^m}}{l}\ddot r_s }+ \right. \left. { Y_l^m{{\ddot r}_s} + \ddot \eta {{\left( {Y_l^m} \right)}^2}} \right). 
\end{align} 
The procedure advances exactly in the same way for the gradient components, giving
\begin{align}
		{\left. {{\partial _\xi }\varphi _{{{\bm{v}}_1}}^{{\text{in}}}} \right|_{\xi = \eta Y_l^m + {r_s}}} &= {\left. {{\xi ^{l - 1}}Y_l^mr_s^{ - l}\left( {\dot \eta {r_s} + 2\eta {{\dot r}_s}} \right) + \frac{{r_s^2{{\dot r}_s}}}{{{\xi ^2}}}} \right|_{\xi = \eta Y_l^m + {r_s}}} \nonumber\\ \label{Eq16} 
		&= Y_l^mr_s^{ - l}\left( {\dot \eta {r_s} + 2\eta {{\dot r}_s}} \right){\left( {{r_s} + \eta Y_l^m} \right)^{l - 1}} + \frac{{r_s^2{{\dot r}_s}}}{{{{\left( {\eta Y_l^m + {r_s}} \right)}^2}}} \nonumber\\ 
		&\simeq \dot \eta Y_l^m + {{\dot r}_s}, 
\end{align}
\begin{align}		
		{\left. {\frac{{{\partial _\theta }\varphi _{{{\bm{v}}_1}}^{{\text{in}}}}}{\xi }} \right|_{\xi = \eta Y_l^m + {r_s}}} &= {\left. {\frac{{{\xi ^{l - 1}}}}{l}r_s^{ - l}\left( {\dot \eta {r_s} + 2\eta {{\dot r}_s}} \right){\partial _\theta }Y_l^m} \right|_{\xi = \eta Y_l^m + {r_s}}} \nonumber\\ \label{Eq17} 
		&= \frac{{r_s^{ - l}}}{l}\left( {\dot \eta {r_s} + 2\eta {{\dot r}_s}} \right){\left( {\eta Y_l^m + {r_s}} \right)^{l - 1}}{\partial _\theta }Y_l^m \nonumber\\ 
		&\simeq \frac{{2\eta {{\dot r}_s}}}{{l{r_s}}}{\partial _\theta }Y_l^m + \frac{{\dot \eta }}{l}{\partial _\theta }Y_l^m,   
\end{align}
and
\begin{align}			
		{\left. {\frac{{{\partial _\phi }\varphi _{{{\bm{v}}_1}}^{{\text{in}}}}}{{\xi \sin \theta }}} \right|_{\xi = \eta Y_l^m + {r_s}}} &= {\left. {\frac{{{\xi ^{l - 1}}r_s^{ - l}}}{l}\left( {\dot \eta {r_s} + 2\eta {{\dot r}_s}} \right)\csc \theta {\partial _\phi }Y_l^m} \right|_{\xi = \eta Y_l^m + {r_s}}} \nonumber\\ \label{Eq18} 
		&= \frac{{r_s^{ - l}}}{l}\left( {\dot \eta {r_s} + 2\eta {{\dot r}_s}} \right){\left( {{r_s} + \eta Y_l^m} \right)^{l - 1}}\csc \theta {\partial _\phi }Y_l^m \nonumber\\ 
		&\simeq \frac{{2\eta {{\dot r}_s}}}{{l{r_s}}}\csc \theta {\partial _\phi }Y_l^m + \frac{{\dot \eta }}{l}\csc \theta {\partial _\phi }Y_l^m.  
\end{align}
We preferred a slightly longer formalism in the first lines of these equations to show the terms proportional to $ \xi $ so that in the second lines we simplify their substitution at the perturbed location, and in the third lines (Eqs.\eqref{Eq16}, \eqref{Eq17}, \eqref{Eq18}) the remaining terms emerge more clearly. Other more compact formulas can be worked out if necessary but reduce the readability. Eq.\eqref{Eq13} to the first order on the perturbation is then:
\begin{equation}\label{Eq19}
	\frac{p}{{{\rho _{{\text{in}}}}}} + {a_{O'}}\cos \vartheta \left( {\eta Y_l^m + {r_s} - 1} \right) + \ddot \eta {r_s}\frac{{Y_l^m}}{l} + 2\eta {\ddot r_s}\frac{{Y_l^m}}{l} + \eta Y_l^m{\ddot r_s} + 3\dot \eta {\dot r_s}\frac{{Y_l^m}}{l} - {r_s}{\ddot r_s} - \frac{{3\dot r_s^2}}{2} + {\Phi _{\bm{g}}} = 0.
\end{equation}
From this equation we can calculate the pressure. As a ``sanity check'', if we require the reference system to be inertial, then the apparent forces disappear ${a_{O'}} = 0$ and for a zero flow velocity as well as for the case of no perturbation $\eta  = 0$ we get $\frac{p}{{{\rho _{{\text{in}}}}}} + {\Phi _{\bm{g}}} = {r_s}{\ddot r_s} + \frac{3}{2}\dot r_s^2$ which is the standard literature equation of the expanding/contracting bubble for zero surface tension \citep[e.g.,][]{2000ifd..book.....B}. Finally, it is evident that when the perturbation is not null but no velocity fluid is included $v = 0$ we obtain the results in \citet[][]{1954JAP....25...96P}. Hence in these cases our results reduce to to well-known results in the literature. 

\subsubsection{External gas pressure equation}\label{Externalgaspressureequation}
In the external gas case, the pressure equation is again obtained from the Bernoulli equation by adding the inertial term as in the previous section. Here we pay for writing our equations in ${S_1}$ instead of ${S_0}$ with a slightly more complex formalism; nevertheless, the procedure is the same as the one outlined above and the approach will  result in an easier physical interpretation of our final results. We consider the terms in the following equation
\begin{equation}\label{Eq20}
	{\partial _t}\varphi _{{{\bm{v}}_1}}^{{\text{out}}} + \frac{1}{2}\left\langle {{\nabla _{\bm{\xi }}}\varphi _{{{\bm{v}}_1}}^{{\text{out}}},{\nabla _{\bm{\xi }}}\varphi _{{{\bm{v}}_1}}^{{\text{out}}}} \right\rangle  + \frac{p}{{{\rho _{{\text{out}}}}}} = \frac{{v_{{\text{rel}}}^2}}{2} - {\Phi _{\bm{g}}} - \left\langle {{{\bm{a}}_{O'}},{\bm{\xi }}} \right\rangle ,
\end{equation}
which we evaluate at the perturbed location $\xi  = \eta Y_l^m + {r_s}$. Here the free function ${f^{{\text{out}}}} = \frac{{v_{{\text{rel}}}^2}}{2}$ has been previously derived in Paper I (their eq.8) and ${v_{{\text{rel}}}}$ is the velocity of the fluid impacting the stellar system in ${S_1}$, i.e. the velocity of the stellar system itself (apart from the sign). The reason for calling it now  ${v_{{\text{rel}}}}$, instead of simply $\left\| {{{\bm{v}}_{O'}}} \right\| = v$, will be  clearer later on.

For each term in Eq.\eqref{Eq20} to the first order we obtain:
\begin{align}
  {\left. {{\partial _t}\varphi _{{{\bm{v}}_1}}^{{\text{out}}}} \right|_{\xi = \eta Y_l^m + {r_s}}} &\simeq  - \frac{{3\eta {a_{O'}}\sin \vartheta {\partial _\theta }Y_l^m}}{{2(l + 1)}} - \frac{{3{v_{{\text{rel}}}}\sin \theta \left( {\eta (l + 1){{\dot r}_s} + \dot \eta {r_s}} \right){\partial _\theta }Y_l^m}}{{2(l + 1){r_s}}} - {a_{O'}}\cos \vartheta \left( {\frac{{3\eta Y_l^m}}{{l + 1}} + \frac{{3{r_s}}}{2}} \right) \nonumber\\ \label{Eq21}
   &- {v_{{\text{rel}}}}\cos \theta \left( {\frac{{3\dot \eta Y_l^m}}{{l + 1}} + \frac{{3{{\dot r}_s}}}{2}} \right) - \frac{{\ddot \eta {r_s}Y_l^m}}{{l + 1}} - {{\ddot r}_s}{r_s} + \eta {{\ddot r}_s}\frac{{l - 1}}{{l + 1}}Y_l^m - \dot \eta \frac{{l + 4}}{{l + 1}}{{\dot r}_s}Y_l^m - 2\frac{{l - 1}}{{l + 1}}\dot r_s^2, \\ \label{Eq22}
  {\left. {{\partial _\xi }\varphi _{{{\bm{v}}_1}}^{{\text{out}}}} \right|_{\xi = \eta Y_l^m + {r_s}}} &\simeq \frac{{3\eta }}{{2{r_s}}}{v_{{\text{rel}}}}\sin \theta {\partial _\theta }Y_l^m + \dot \eta Y_l^m + {{\dot r}_s}, \\ \label{Eq23}
  {\left. {\frac{{{\partial _\theta }\varphi _{{{\bm{v}}_1}}^{{\text{out}}}}}{\xi }} \right|_{\xi = \eta Y_l^m + {r_s}}} &\simeq \frac{{3{v_{{\text{rel}}}}\sin \theta }}{{2(l + 1){r_s}}}\left( {(l + 1){r_s} - \eta \left( {{\partial _{\theta ,\theta }}Y_l^m + (l - 1)Y_l^m} \right)} \right) - \frac{{9\eta {\partial _\theta }Y_l^m}}{{2(l + 1){r_s}}}{v_{{\text{rel}}}}\cos \theta  - \frac{{\dot \eta {r_s} + 2\eta {{\dot r}_s}}}{{(l + 1){r_s}}}{\partial _\theta }Y_l^m, \\ \label{Eq24}
  {\left. {\frac{{{\partial _\phi }\varphi _{{{\bm{v}}_1}}^{{\text{out}}}}}{{\xi \sin \theta }}} \right|_{\xi = \eta Y_l^m + {r_s}}} &\simeq  - \frac{{3\eta v_{\text{rel}}{\partial _\phi }Y_l^m}}{{(l + 1){r_s}}}\cot \theta  - \frac{{3\eta {v_{{\text{rel}}}}{\partial _{\theta ,\phi }}Y_l^m}}{{2(l + 1){r_s}}} - \frac{{\csc \theta \left( {\dot \eta {r_s} + 2\eta {{\dot r}_s}} \right)}}{{(l + 1){r_s}}}{\partial _\phi }Y_l^m.  
\end{align} 
Eq.\eqref{Eq20} is simplified by collecting Eqs.\eqref{Eq21}, \eqref{Eq22}, \eqref{Eq23} and \eqref{Eq24} once the scalar product is taken into account. As before, the solution of Eq.\eqref{Eq20} can be obtained in terms of the pressure $p$ and it can be simplified by retaining only the first order terms. We have (hereafter we define ${l_ + } \equiv l + 1$, ${l_{ +  + }} \equiv l + 2$ and ${l_ - } \equiv l - 1$ etc. to minimize the notation) 
\begin{equation}\label{Eq25}
	\begin{gathered}
\frac{p}{{{\rho _{{\rm{out}}}}}} - \frac{3}{4}\frac{{{\partial _\theta }Y_l^m\sin \vartheta }}{{{l_ + }}}\eta {a_{O'}} + v_{{\rm{rel}}}^2\left( {\frac{5}{8} - \frac{9}{4}\frac{\eta }{{{r_s}}}\frac{{{\partial _{\theta ,\theta }}Y_l^m + {l_ - }Y_l^m}}{{{l_ + }}}} \right){\sin ^2}\theta  - 3{v_{{\rm{rel}}}}\left( {\dot \eta {r_s} + \eta {{\dot r}_s}} \right)\frac{{{\partial _\theta }Y_l^m}}{{{l_ + }{r_s}}}\sin \theta \\
 - \frac{9}{4}\frac{{3\eta v_{{\rm{rel}}}^2}}{{{r_s}}}\frac{{{\partial _\theta }Y_l^m}}{{{l_ + }}}\sin \theta \cos \theta  + {a_{O'}}\left( {\eta \frac{{{l_{ -  - }}Y_l^m}}{{{l_ + }}} - \frac{{{r_s}}}{2}} \right)\cos \vartheta  + {v_{{\rm{rel}}}}\left( { - \frac{{3\dot \eta Y_l^m}}{{{l_ + }}} - \frac{{3{{\dot r}_s}}}{2}} \right)\cos \theta \\
 - \ddot \eta {r_s}{l_ + } - {{\ddot r}_s}\left( {{r_s} - \eta \frac{{{l_ - }Y_l^m}}{{{l_ + }}}} \right) - 3\dot \eta {{\dot r}_s}\frac{{Y_l^m}}{{{l_ + }}} - 3\dot r_s^2 - \frac{1}{2}{\cos ^2}\theta v_{{\rm{rel}}}^2 + {\Phi _{\bm{g}}} = 0. \hfill \\ 
	\end{gathered}
\end{equation}
Again we can check the validity of this equation by assuming no perturbation $\eta  \to 0$ and $l = 0$ to prove that it effectively reduces to the Theorem of Section 3 in \citet{2014arXiv1403.6122P} as a particular case. 

\subsubsection{Surface of equilibrium}\label{Surfaceofequilibrium}
Taking the difference between Eq.\eqref{Eq19} and \eqref{Eq25}, we express the continuity condition of the pressure impacting on the stellar system from the external gas (the ram pressure condition). 
The equation of motion for the unperturbed equation is:
\begin{equation}\label{Eq26}
	{v_\text{rel}^2}{\cos ^2}\theta  - \frac{{{\rho _{{\rm{in}}}} - \frac{5}{4}{\rho _{{\rm{out}}}}}}{{{\rho _{{\rm{out}}}} - {\rho _{{\rm{in}}}}}}{v_\text{rel}^2}{\sin ^2}\theta  + 3\frac{{{\rho _{{\rm{out}}}}}}{{{\rho _{{\rm{out}}}} - {\rho _{{\rm{in}}}}}}{\dot r_s}v_\text{rel}\cos \theta  + {a_{O'}}\cos \vartheta {r_s}\frac{{2{\rho _{{\rm{in}}}} + {\rho _{{\rm{out}}}}}}{{{\rho _{{\rm{out}}}} - {\rho _{{\rm{in}}}}}} + \left( {2{\Phi _{\bm{g}}} - 2{r_s}{{\ddot r}_s} - 3\dot r_s^2} \right) = 0.
\end{equation}
Our disposition of the terms indicates immediately that in ${S_0}$, without motion of the fluid or the sphere, we obtain $\left( {{\rho _{{\rm{in}}}} - {\rho _{{\rm{out}}}}} \right)\left( {2{\Phi _{\bm{g}}} - 2{r_s}{{\ddot r}_s} - 3\dot r_s^2} \right) = 0$, which indicates the condition of equilibrium where always ${\rho _{{\rm{in}}}} \ne {\rho _{{\rm{out}}}}$ as ${\Phi _{\bm{g}}} = {r_s}{\ddot r_s} + \frac{3}{2}\dot r_s^2$. This, for example, may describe a case of a galaxy lying at the centre of a galaxy cluster.
Therefore, because we are interested in the growth of the perturbation over an equilibrium state (for at least one instability mode), in the resulting equation we need to study only the terms proportional to the perturbation terms (i.e., the terms containing the spherical harmonics) that we analyse in the next section. 
 We move from this equation in order to investigate the more interesting case of the differential equation for $\eta  = \eta \left( t \right)$ from which, stability condition for the growth of a perturbation can be derived. 

\subsection{Condition for the instability}\label{Conditionfortheinstability}
The condition for the instability is derived by considering only the perturbed terms in difference between Eq.\eqref{Eq19} and \eqref{Eq25}. Collecting terms in $Y_l^m$ and its derivatives we obtain an equation of the form ${a_1}Y_l^m + {a_2}{\partial _\theta }Y_l^m + {a_3}{\partial _{\theta ,\theta }}Y_l^m = 0$ for some form of the functions ${a_i} = {a_i}\left( {\eta ,\dot \eta ,\ddot \eta } \right)$ and $i=1, 2, 3$, which suggests that we define two special functions as follows:
\begin{equation}\label{Eq27}
	{F_1} = {F_1}\left( {\theta ,l,m} \right) \equiv \frac{{{\partial _\theta }Y_l^m}}{{Y_l^m}}
\end{equation}
\begin{equation}\label{Eq28}
	{F_2} = {F_2}\left( {\theta ,l,m} \right) \equiv \frac{{{\partial _{\theta ,\theta }}Y_l^m}}{{Y_l^m}},
\end{equation}
independent from $\eta $ or its derivatives. In this way we obtain an equation for the perturbation $\eta $ of the form ${a_1}\left( {\eta ,\dot \eta ,\ddot \eta } \right) + {a_2}\left( {\eta ,\dot \eta ,\ddot \eta } \right){F_1} + {a_3}\left( {\eta ,\dot \eta ,\ddot \eta } \right){F_2} = 0$ which immediately produces an interesting result as follow: \textit{the presence of a preferential direction for the motion of the galaxy along its orbit induces a symmetry on the perturbations}. The dependence on the considered azimuthal mode remains, i.e. the dependence on $m$, nevertheless it becomes independent from the azimuthal direction $\phi $. This is an interesting simplification that is a consequence of the geometry assumed.

The study of the stability of the solution of an equation of the form ${a_1}\left( {\eta ,\dot \eta ,\ddot \eta } \right) + {a_2}\left( {\eta ,\dot \eta ,\ddot \eta } \right){F_1} + {a_3}\left( {\eta ,\dot \eta ,\ddot \eta } \right){F_2} = 0$ is better performed if we convert it to an eigen-value problem. To proceed in this way we collect the terms depending on the perturbation factor $\eta$ and its derivatives. With the aid of Eq.\eqref{Eq27} and \eqref{Eq28} we put the differential equation  ${a_1}\left( {\eta ,\dot \eta ,\ddot \eta } \right) + {a_2}\left( {\eta ,\dot \eta ,\ddot \eta } \right){F_1} + {a_3}\left( {\eta ,\dot \eta ,\ddot \eta } \right){F_2} = 0$ in standard form. Hence, the more suitable form for starting our stability analysis obtained by taking only the perturbed terms that differ between Eq.\eqref{Eq19} and \eqref{Eq25} and accounting for Eqs.\eqref{Eq27} and \eqref{Eq28} is
\begin{align}\label{Eq30}
&\ddot \eta  + \frac{{3{v_{{\rm{rel}}}}}}{{{r_s}}}\left( {\frac{{l{\rho _{{\rm{out}}}}\left( {{S_1}\sin \theta  + \cos \theta } \right)}}{{{l_ + }{\rho ^{{\rm{in}}}} + l{\rho ^{{\rm{out}}}}}} + {{\dot r}_s}} \right)\dot \eta  + \left( {\frac{{l{A_{O'}}\cos \vartheta }}{{{r_s}}}\frac{{{l_ + }{\rho _{{\rm{in}}}} - {l_{ -  - }}{\rho _{{\rm{out}}}}}}{{{l_ + }{\rho _{{\rm{in}}}} + l{\rho _{{\rm{out}}}}}}} \right.\nonumber\\
&\left. { + \frac{3}{2}\frac{{l{S_1}{A_{O'}}\sin \vartheta }}{{{r_s}}}\frac{{{\rho _{{\rm{out}}}}}}{{{l_ + }{\rho _{{\rm{in}}}} + l{\rho _{{\rm{out}}}}}} + \frac{{v\sin \theta }}{{r_s^2}}\frac{{l{\rho _{{\rm{out}}}}}}{{{l_ + }{\rho _{{\rm{in}}}} + l{\rho _{{\rm{out}}}}}}\left( {\frac{9}{4}v\sin \theta ({l_ - } + {S_2}) + 3{S_1}\left( {{{\dot r}_s} + \frac{9}{4}v\cos \theta } \right)} \right) + {\rm A}} \right)\eta  = 0.
\end{align}
Despite its complicated form, this equation is formulated in a suitable way to show that it can reduce to Eq.(13) in \citet{1954JAP....25...96P}.

Considering we have no known terms in the left hand side (LHS) of Eq.\eqref{Eq30}, i.e. it is a second order ODE of the type $\ddot \eta  + a\left( t \right)\dot \eta  + b\left( t \right)\eta  = 0$ for $\eta  = \eta \left( t \right)$, we can attempt a classical quantum mechanics Wentzel$-$Kramers$-$Brillouin (WKB) approach to the solution by making use of the transformation 
\begin{equation}\label{Eq31}
\eta   \left( t \right) = \alpha \left( t \right){e^{ - \frac{1}{2}\int_{{t_0}}^t {\Theta\left( \tau  \right)d\tau } }},
\end{equation}
where on purpose we choose,
$$
\Theta \left( \tau  \right) \equiv \frac{{3{v_{{\rm{rel}}}}}}{{{r_s}}}\left( {{{\dot r}_s} + \frac{{l{\rho _{{\rm{out}}}}}}{{{l_ + }{\rho ^{{\rm{in}}}} + l{\rho ^{{\rm{out}}}}}}\left( {{F_1}\sin \theta  + \cos \theta } \right)} \right),
$$
to simplify Eq.\eqref{Eq30} to a standard eigenvalues problem with slowly varying coefficient:
\begin{equation}\label{Eq32}
	\ddot \alpha  = {\gamma ^2}\left( {\theta ;t} \right)\alpha,
\end{equation}
whose solution in conveniently carried out in WKB approximation. However, we will accomplish a much simpler task here. We are interested in the condition for which at least one mode is unstable, and the instability of the harmonic oscillator equation Eq.\eqref{Eq32} is well known to depend on the positivity of the growth factor ${\gamma ^2}\left( {\theta ;t} \right) > 0$ where
\begin{align}\label{ApEq33}
{\gamma ^2}\left( {\theta ;t} \right) &\equiv  - \frac{{{a_{O'}}\cos \vartheta }}{{2{r_s}}}\frac{{l\left( {2{l_ + }{\rho _{{\rm{in}}}} - \left( {l + {l_ - }} \right){\rho _{{\rm{out}}}}} \right)}}{{{l_ + }{\rho _{{\rm{in}}}} + l{\rho _{{\rm{out}}}}}} + \frac{9}{4}\frac{{v_{{\rm{rel}}}^2{{\cos }^2}\theta }}{{r_s^2}}{\left( {\frac{{l{\rho _{{\rm{out}}}}}}{{{l_ + }{\rho _{{\rm{in}}}} + l{\rho _{{\rm{out}}}}}}} \right)^2} - \frac{9}{4}\frac{{v_{{\rm{rel}}}^2\sin \theta \cos \theta }}{{r_s^2}}\frac{{{F_1}l{\rho _{{\rm{out}}}}\left( {3{l_ + }{\rho _{{\rm{in}}}} + l{\rho _{{\rm{out}}}}} \right)}}{{{{\left( {{l_ + }{\rho _{{\rm{in}}}} + l{\rho _{{\rm{out}}}}} \right)}^2}}}\nonumber\\
 &+ \frac{{{{\dot r}_s}{v_{{\rm{rel}}}}\cos \theta }}{{r_s^2}}\frac{{3l{\rho _{{\rm{out}}}}}}{{{l_ + }{\rho _{{\rm{in}}}} + l{\rho _{{\rm{out}}}}}} + \frac{9}{4}\frac{{v_{{\rm{rel}}}^2{{\sin }^2}\theta }}{{r_s^2}}\frac{{{{\left( {{F_1}l{\rho _{{\rm{out}}}}} \right)}^2} - \left( {{l_ - } + {F_2}} \right)l{\rho _{{\rm{out}}}}\left( {{l_ + }{\rho _{{\rm{in}}}} + l{\rho _{{\rm{out}}}}} \right)}}{{{{\left( {{l_ + }{\rho _{{\rm{in}}}} + l{\rho _{{\rm{out}}}}} \right)}^2}}}\nonumber\\
 &+ \frac{3}{4}\frac{{\dot r_s^2}}{{r_s^2}} - \frac{{(l + {l_ + })\left( {{l_ + }{\rho _{{\rm{in}}}} - l{\rho _{{\rm{out}}}}} \right)}}{{2\left( {{l_ + }{\rho _{{\rm{in}}}} + l{\rho _{{\rm{out}}}}} \right)}}\frac{{{{\ddot r}_s}}}{{{r_s}}}.
\end{align}
This represents the desired equation already presented in a more compact fashion in Eq.\eqref{Eq37}. For the purpose of this Appendix of recovering some limit-cases we will explicitly keep the terms in Eq. \eqref{ApEq33} written in their full extension.  ${\gamma ^2}$ relates to the growth of the perturbation for which we were searching (it is indeed called the growth factor). This completes our theoretical framework and equips us with the tools to investigate the growth of the instabilities by compression or instabilities that lead to star formation.

Before embarking onto the analysis of the growth factor, we recover some classical literature limits to validate the physics we will encounter. 

\subsubsection{Special limits}\label{SpLims}
We start by remarking how in the case of a non-inertial reference frame ${S_1}$, the instability condition reduces to the study of the positivity of the last row (i.e. the third) of Eq.\eqref{ApEq33} (that we rewrite in a compact way as (with Eq. \eqref{Eq29} of Section \ref{GeometricalSoR}):
\begin{equation}\label{Eq34}
\left( {l + \frac{1}{2}} \right){\rm A} + \frac{{3\dot r_s^2}}{{4r_s^2}} > 0.
\end{equation}
This equation was already presented by \citet[][]{1954JAP....25...96P} (his Eq.(17) with zero surface tension).

Another important limit to recover is the plane case. In the spherical geometry that we have assumed, the plane case can be achieved by taking $l \to \infty$ and $R \to \infty $  and keeping the wave number of the perturbation, $k$,  constant. This is not a trivial task for the presence of the special functions ${F_1}$  and  ${F_2}$ defined in Eq.\eqref{Eq27} and \eqref{Eq28} whose dependence on $l$ involves the determination of the Euler Gamma function for large values of the index $l$. We refer the interested reader to Appendix B for the computation of their asymptotic behaviour for large $l$ because of its exclusively mathematical nature. Using the results of Appendix B we can show that Eq.\eqref{ApEq33} behaves in the plane case as 
\begin{equation}\label{Eq35}
	\gamma _{{\rm{plane}}}^2 \simeq \frac{{k{a_{O'}}\cos \vartheta \left( {{\rho _{{\rm{out}}}} - {\rho _{{\rm{in}}}}} \right)}}{{{\rho _{{\rm{in}}}} + {\rho _{{\rm{out}}}}}} - \frac{9}{4}\frac{{{k^2}{{\sin }^2}\theta {\rho _{{\rm{out}}}}v_{{\rm{rel}}}^2}}{{{\rho _{{\rm{in}}}} + {\rho _{{\rm{out}}}}}} - \frac{{k\left( {{\rho _{{\rm{out}}}} - {\rho _{{\rm{in}}}}} \right){{\ddot r}_s}}}{{{\rho _{{\rm{in}}}} + {\rho _{{\rm{out}}}}}}.
\end{equation}
Then, if we define, as usual, the acceleration to be ${\ddot r_s} = g$ the previous equation reduces to 
\begin{equation}\label{Eq36}
 \gamma _{{\rm{plane}}}^2 \simeq \frac{{k\left( {{\rho _{{\rm{out}}}} - {\rho _{{\rm{in}}}}} \right)\left( {{a_ \bot } - g} \right)}}{{{\rho _{{\rm{in}}}} + {\rho _{{\rm{out}}}}}} - \frac{9}{4}\frac{{{k^2}{{\sin }^2}\theta {\rho _{{\rm{out}}}}\left( {{v_{{\rm{in}}}} - {v_{{\rm{out}}}}} \right)}}{{{\rho _{{\rm{in}}}} + {\rho _{{\rm{out}}}}}}, 
\end{equation}
which can  be easily interpreted remembering the plane case in the literature as discussed for Eq.\eqref{Eq01}. We indicated with ${a_ \bot }$ the acceleration orthogonal to the surface that in the plane case represents the vertical direction. Hence, the first term is exactly the instability criterion for the RT effect where the effective acceleration ${g_{{\rm{eff}}}} = {a_ \bot } - g$ has been corrected for the presence of the corrective-term ${a_ \bot }$. In the same way, the second term retains the key dependencies from the relative velocity ${v_{{\rm{rel}}}} = {v_{{\rm{in}}}} - {v_{{\rm{out}}}}$ between the fluid above and below the surface dividing the two sliding fluids that are the basis of the KH instability. These criteria become equivalent to the RT and KH criteria (apart from the numerical factors $9/4{\rho _{{\rm{out}}}}$) at the stagnation point, where $\cos \vartheta  = 1$ and ${\sin ^2}\theta  = 0$ and at the tangent to the sphere $\cos \vartheta  = 0$ and ${\sin ^2}\theta  = 1$ respectively.

\section{Asymptoctic expansion of the functions ${F_1}$  and ${F_2}$}\label{SpecialFunc}
We elaborate in this appendix on more mathematical theorems that can be skipped in a first reading.

\begin{figure*}
\includegraphics[width=\columnwidth]{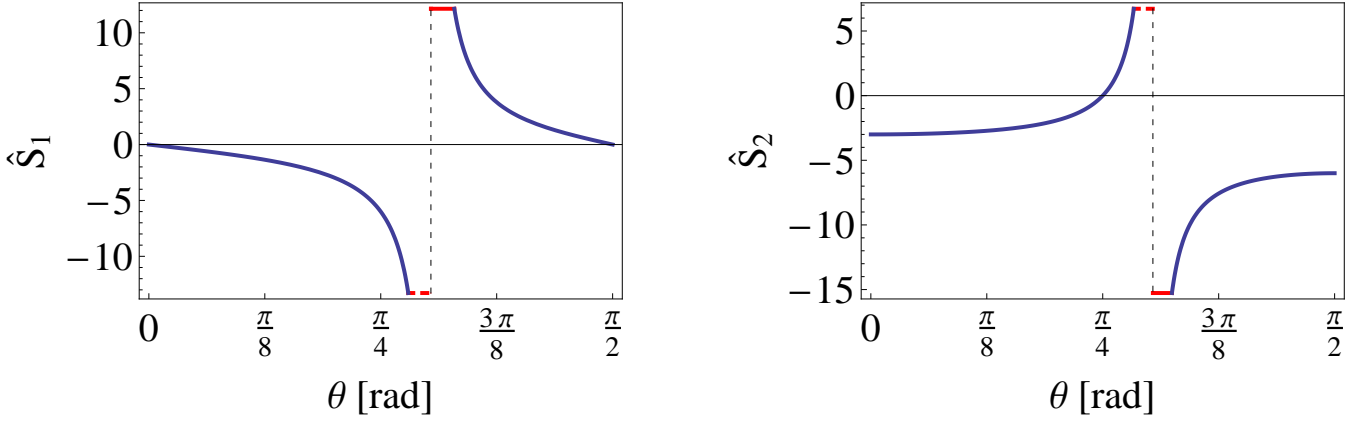}
\caption{Special function $F_1$ and $F_2$ for the modal perturbation $l=2$. The divergence (dashed vertical line) is located at $\left( {\theta  = \frac{1}{2}\left( {2\pi c - {{\cos }^{ - 1}}\left( { - \frac{1}{3}} \right)} \right) \vee \theta  = \frac{1}{2}\left( {2\pi c + {{\cos }^{ - 1}}\left( { - \frac{1}{3}} \right)} \right)} \right) \wedge c \in \mathbb{Z}$. Only the angular range of interest $\theta  \in \left[ {0,\frac{\pi }{2}} \right]$ is accounted.
\label{FigS1l2} }
\end{figure*}

We are interested in the limits of the special functions ${F_1}$  and ${F_2}$ defined and in their asymptotic expansion. A plot of the two functions for the instability mode of interest ($l=2$) and the angular dependence of interest $\theta  \in \left[ {0,\frac{\pi }{2}} \right]$, is presented in Fig.\ref{FigS1l2}, where ${\hat F_1} = {F_1}\left( {\theta ,2,0} \right) =  - \frac{{6\sin (2\theta )}}{{3\cos (2\theta ) + 1}} \simeq  - 3\theta  + O{\left( \theta  \right)^2}$ and ${\hat F_2} = {F_2}\left( {\theta ,2,0} \right) = \frac{4}{{3\cos (2\theta ) + 1}} - 4 \simeq  - 3 + O{\left( \theta  \right)^2}$ respectively.

As the index $l$ tends to approach $\infty $ we can write for $\theta  \in \left[ { - \frac{\pi }{2},\frac{\pi }{2}} \right]$ 
\begin{align}
{F_1} &\equiv \frac{1}{{Y_l^m}}\frac{{\partial Y_l^m}}{{\partial \theta }}\\
 &= \frac{{P_l^{m + 1}(\cos (\theta ))}}{{P_l^m(\cos (\theta ))}} + m\cot (\theta ),
\end{align}
while
\begin{align}
{F_2} &\equiv \frac{1}{{Y_l^m}}\frac{{{\partial ^2}Y_l^m}}{{\partial {\theta ^2}}}\\
 &= \frac{{P_l^{m + 2}\left( {\cos \theta } \right) + (2m + 1)\cot \theta P_l^{m + 1}\left( {\cos \theta } \right)}}{{P_l^m\left( {\cos \theta } \right)}} + m\left( {m{{\cot }^2}\theta  - {{\csc }^2}\theta } \right).
\end{align}
As already discussed in association with Fig.\ref{TTEvol} there is no observational evidence of strong azimuthal asymmetries in dwarf galaxies of the Local Group, thus it is safe to assume the perturbation to be well represented by modes with $m = 0$. With this assumption, in the stagnation point direction $\theta  = 0$ is 
\begin{align}
{F_1}\left( {0,\infty ,0} \right) &= \mathop {\lim }\limits_{l \to \infty } \left( {\mathop {\lim }\limits_{\theta  \to 0} {{\left. {\frac{1}{{Y_l^m}}\frac{{\partial Y_l^m}}{{\partial \theta }}} \right|}_{m = 0}}} \right)\\
 &= \mathop {\lim }\limits_{l \to \infty } \frac{{P_l^1\left( 1 \right)}}{{P_l^0\left( 1 \right)}}\\
 &= 0.
\end{align}
More cumbersome is the same limit for the case $\theta  = \frac{\pi }{2}$. We get
\begin{align}
{F_1}\left( {\frac{\pi }{2},\infty ,0} \right) &= \mathop {\lim }\limits_{l \to \infty } \left( {\mathop {\lim }\limits_{\theta  \to \frac{\pi }{2}} {{\left. {\frac{1}{{Y_l^m}}\frac{{\partial Y_l^m}}{{\partial \theta }}} \right|}_{m = 0}}} \right)\\
 &= 2\mathop {\lim }\limits_{l \to \infty } \frac{{\Gamma \left( {\frac{1}{2} - \frac{l}{2}} \right)\Gamma \left( {1 + \frac{l}{2}} \right)}}{{\Gamma \left( {\frac{1}{2} + \frac{l}{2}} \right)\Gamma \left( {0 - \frac{l}{2}} \right)}}\\
 &= 2\mathop {\lim }\limits_{n \to \infty } \frac{{\Gamma \left( {\frac{1}{2} - n} \right)\Gamma (n + 1)}}{{\Gamma (n)\Gamma \left( {n + \frac{1}{2}} \right)}}\\
 &\simeq \mathop {\lim }\limits_{n \to \infty } 2{e^{n\left( {2 + \log \left( {\frac{1}{n}} \right)} \right) + O{{\left( {\frac{1}{n}} \right)}^2}}}\sec \left( {n\pi } \right)\left( {\frac{n}{2} + \frac{1}{{24}} + O{{\left( {\frac{1}{n}} \right)}^2}} \right)\\
 &= 0,
\end{align}
where in order to prove this theorem the Stirling expansion for the Gamma function has to be considered.

By analogy with the previous proofs we get
\begin{align}
{F_2}\left( {0,\infty ,0} \right) &= \mathop {\lim }\limits_{l \to \infty } \left( {\mathop {\lim }\limits_{\theta  \to 0} {{\left. {\frac{1}{{Y_l^m}}\frac{{{\partial ^2}Y_l^m}}{{\partial {\theta ^2}}}} \right|}_{m = 0}}} \right)\\
 &= O\left( { - \frac{{l(l + 1)}}{2}} \right),
\end{align}
where we used the ``big-O'' to express that the limit is increasing to infinity as the power written. Once introduced in Eq.\eqref{ApEq33} this behaviour cancels out to the desired limit offering the finite limit we wrote in Eq. \eqref{Eq35}. With the same abuse of notation now clearly we can write
\begin{align}
{F_2}\left( {\frac{\pi }{2},\infty ,0} \right) &= \mathop {\lim }\limits_{l \to \infty } \left( {\mathop {\lim }\limits_{\theta  \to \frac{\pi }{2}} {{\left. {\frac{1}{{Y_l^m}}\frac{{{\partial ^2}Y_l^m}}{{\partial {\theta ^2}}}} \right|}_{m = 0}}} \right)\\
 &= O\left( {\frac{{(l - 1)l(l + 1)(l + 2)\Gamma \left( {\frac{1}{2} - \frac{l}{2}} \right)\Gamma \left( {\frac{l}{2} + 1} \right)}}{{4\Gamma \left( {\frac{3}{2} - \frac{l}{2}} \right)\Gamma \left( {\frac{l}{2} + 2} \right)}}} \right)\\
 &= O\left( { - l\left( {l + 1} \right)} \right).
\end{align}
This proves the asymptotic behaviour at the leading order of the special functions ${F_1}$  and ${F_2}$.


\bibliographystyle{aa} 
\bibliography{BiblioArt} 
\end{document}